%% file: arxiv_v0.tex
\documentclass[11pt]{article}

\usepackage{epsfig}
\usepackage{amsmath, amsthm, amssymb}
\usepackage[round]{natbib}
\usepackage{graphics}
\usepackage{graphicx,pstricks}
\usepackage{rotating}
\usepackage{moreverb}
\usepackage{subfigure}
\usepackage{hangcaption}
\usepackage{txfonts}

\usepackage{hyperref}
\hypersetup{colorlinks=false}
\usepackage[mathscr]{eucal}

\newtheorem{thm}{Theorem}[section]
\newtheorem{cor}[thm]{Corollary}

\newtheorem{prop}[thm]{Proposition}

\newtheorem{remark}[thm]{Remark}

\newcommand{\bfX}{\textbf{X}}
\newcommand{\tbfX}{\widetilde{\textbf{X}}}
\newcommand{\bfI}{\textbf{I}}
\newcommand{\bfy}{\textbf{y}}
\newcommand{\bfh}{\boldsymbol{\kappa}}

\newcommand{\bftheta}{\boldsymbol{\theta}}
\newcommand{\bfbeta}{\boldsymbol{\beta}}
\newcommand{\bfalpha}{\boldsymbol{\alpha}}
\newcommand{\tbfbeta}{\widetilde{\boldsymbol{\beta}}}
\newcommand{\tbfalpha}{\widetilde{\boldsymbol{\alpha}}}
\newcommand{\tbfeta}{\widetilde{\boldsymbol{\eta}}}

\newcommand{\bftau}{\boldsymbol{\tau}}
\newcommand{\bflam}{\boldsymbol{\lambda}}

\newcommand{\R}{{\sf R}~}
\newcommand{\benum}{\begin{eqnarray}}
\newcommand{\eenum}{\end{eqnarray}}
\newcommand{\be}{\begin{eqnarray*}}
\newcommand{\ee}{\end{eqnarray*}}
\newcommand{\bi}{\begin{itemize}}
\newcommand{\ei}{\end{itemize}}
\newcommand{\Real}{\mathbb{R}}
\newcommand{\obfn}{\xi}
\newcommand{\sobfn}{\xi^{SUR}}

\title{MM Algorithms for Minimizing Nonsmoothly Penalized Objective Functions}
\author{Elizabeth D. Schifano\footnote{Department of Statistical Science,
301 Malott Hall, Cornell University, Ithaca NY 14853 USA}, 
Robert L. Strawderman${}^*$, Martin T. Wells${}^*$}
\date{}
\begin{document}
\maketitle

\begin{abstract}

  The use of regularization, or penalization, has become increasingly
  common in high-dimensional statistical analysis over the past
  decade, where a common goal is to simultaneously select important
  variables and estimate their effects. It has been shown by several
  authors that these goals can be achieved by minimizing some
  parameter-dependent ``goodness of fit'' function (e.g., a negative
  loglikelihood) subject to a penalization that promotes sparsity.
  Penalty functions that are nonsmooth (i.e. not differentiable) at
  the origin have received substantial attention, arguably beginning
  with LASSO \citep{tibs}.

  The current literature tends to focus on specific combinations of
  smooth data fidelity (i.e., goodness-of-fit) and nonsmooth penalty
  functions.  One result of this combined specificity has been a
  proliferation in the number of computational algorithms designed to
  solve fairly narrow classes of optimization problems involving
  objective functions that are not everywhere continuously
  differentiable. In this paper, we propose a general class of
  algorithms for optimizing an extensive variety of nonsmoothly
  penalized objective functions that satisfy certain regularity
  conditions.  The proposed framework utilizes the
  majorization-minimization (MM) algorithm as its core optimization
  engine. The resulting algorithms rely on iterated soft-thresholding,
  implemented componentwise, allowing for fast, stable updating that
  avoids the need for any high-dimensional matrix inversion. We
  establish a local convergence theory for this class of algorithms
  under weaker assumptions than previously considered in the
  statistical literature.  We also demonstrate the exceptional
  effectiveness of new acceleration methods, originally proposed for
  the EM algorithm, in this class of problems.  Simulation results and
  a microarray data example are provided to demonstrate the
  algorithm's capabilities and versatility. \\

\end{abstract}

\section{Introduction}

Variable selection is an important and challenging issue in the
rapidly growing realm of high-dimensional statistical modeling.  In
such cases, it is often of interest to identify a few important
variables in a veritable sea of noise.  Modern methods, increasingly
based on the principle of penalized likelihood estimation applied to
high dimensional regression problems, attempt to achieve this goal
through an adaptive variable selection process that simultaneously
permits estimation of regression effects.  Indeed, the literature on
the penalization of a ``goodness of fit'' function (e.g., negative
loglikelihood), with a penalty singular at the origin, is quickly
becoming vast, proliferating in part due to the consideration of
specific combinations of data fidelity (i.e., goodness-of-fit) and
penalty functions, the associated statistical properties of resulting
estimators, and the development of several combination-specific
optimization algorithms, \citep[e.g.,][]{tibs, zou:alas, en,
  zou:zhang, FanLi, glmpath, glmnet}.

In this paper, we propose a unified optimization framework that
appeals to the Majorization-Minimization (MM) algorithm \citep{optim}
as the primary optimization tool. The resulting class of algorithms is
referred to as MIST, an acronym for Minimization by Iterative Soft
Thresholding.  The MM algorithm has been considered before for solving
specific classes of singularly penalized likelihood estimation
problems \citep[e.g.,][]{daub,hunter:li, zou:li}; to a large extent,
this work is motivated by these ideas. A distinct advantage of the
proposed work is the exceptional versatility of the class of MIST
algorithms, their associated ease of implementation and numerical
stability, and the development of a fixed point convergence theory
that permits weaker assumptions than existing papers in this area.  We
emphasize here the focus of this paper is on the development of a
stable and versatile class of algorithms applicable to a wide variety
of singularly penalized estimation problems. In particular, the
consideration of asymptotic and oracle properties of estimators
derived from particular combinations of fidelity and penalty
functions, as well as methods for effectively choosing associated
penalty parameters, are not focal points of this paper. A
comprehensive treatment of these results may be found in \cite{johnson},
where asymptotics and oracle properties for estimators derived from a
general class of penalized estimating equations are developed in some
detail.

The paper is organized as follows.  In Section 2, we introduce
notation and provide sufficient conditions for local convergence of
the MM algorithm applied to a large class of data-fidelity and
non-smooth penalty functions.  In Section 3, we present a specialized
version of this general algorithm, demonstrating in particular how the
minimization step of the MM algorithm can be carried out using
iterated soft-thresholding. In its most general form, iterated
soft-thresholding is required at each minimization step; we further
demonstrate how to carry out this minimization step in one iteration
through a judicious choice of majorization function.  As a
consequence, we present a simplified class of iterative algorithms
that are applicable to a wide class of singularly penalized,
generalized linear regression models.

Simulation results are provided in Section 4, while an application in
survival analysis to Diffuse Large B Cell Lymphoma expression data
\citep{rosetal} is presented in Section 5.  We conclude with a
discussion in Section 6.  Proofs and other relevant results are
collected in the Appendix.

\section{MM algorithms for nonsmooth objective functions}
\label{sec:conv}

Let $\obfn(\bfbeta)$ denote a real-valued objective function to be
minimized for $\bfbeta = (\beta_1,\ldots, \beta_p)^T$ in some convex
subset ${\cal B}$ of $\Real^p$.  Let $\sobfn(\bfbeta,\bfalpha)$ denote
a real-valued ``surrogate'' objective function, where $\bfalpha \in
{\cal B}$. Define the minimization map
\benum
\label{the map}
M(\bfalpha) = \mbox{arg} \!\!\!\!
\min_{\bfbeta \in {\cal B}~~~~}\sobfn( \bfbeta, \bfalpha).
\eenum
Then, if $\sobfn(\bfbeta,\bfalpha)$ majorizes $\obfn(\bfbeta)$ for
each $\bfalpha$, a generic MM algorithm for minimizing
$\obfn(\bfbeta)$ takes the following form \citep[e.g.,][]{optim}:
\begin{enumerate}
\item Initialize $\bfbeta^{(0)}$
\item For $n \geq 0$, compute $\bfbeta^{(n+1)} = M(\bfbeta^{(n)}),$
iterating until convergence.
\end{enumerate}
Provided that the objective function, its surrogate and the mapping
$M(\cdot)$ satisfy certain regularity conditions, one can establish
convergence of this algorithm to a local or global solution.
\citet[][Ch.\ 10]{optim} develops such a theory assuming that the
objective functions $\obfn(\bfbeta)$ and $\sobfn(\bfbeta,\bfalpha)$
are twice continuously differentiable. For problems that lack this
degree of smoothness (e.g., all singularly penalized regression
problems, including lasso, \cite{tibs}; adaptive lasso,
\cite{zou:alas}; and SCAD, \cite{FanLi}), a more general theory of
local convergence is required. One such theory is summarized in
Appendix \ref{app:MMlocal}; related results for the EM algorithm may
be found in \cite{Wu}, \cite{tseng} and \cite{ch08}.

Let $\| \cdot \|$ denote the usual Euclidean vector norm.  Based on
the theory summarized in Appendix \ref{app:MMlocal}, we propose a new and
general class of algorithms for minimizing penalized objective
functions of the form
\benum
\label{the objective}
\obfn(\bfbeta) = g(\bfbeta) + p(\bfbeta;\bflam) + \lambda \varepsilon
\| \bfbeta \|^2,
~~ \lambda > 0, ~\varepsilon \geq 0
\eenum
where $g(\bfbeta)$ and $p(\bfbeta;\bflam)$ are respectively data
fidelity (e.g., negative loglikelihood) and penalty functions that
satisfy regularity conditions to be delineated below.  As will be shown
later, the class of problems represented by \eqref{the objective}
contains all of the penalized regression problems commonly considered
in the current literature. It also covers numerous other problems by
expanding the class of permissible fidelity and penalty functions in a
 substantial way.

We assume throughout that $g(\bfbeta)$ is
convex and coercive for $\bfbeta \in {\cal B}$, where
${\cal B}$ is an open convex subset of $\Real^p$.
We further assume that
\benum
\label{eq:gen_pen}
p(\bfbeta; \bflam) = \sum_{j=1}^p \tilde{p}(|\beta_j|;\bflam_j),
\eenum where the vector $\bflam = (\bflam_1^T,\ldots, \bflam_p^T)^T$
and $\bflam_j$ denotes the block of $\bflam$ associated with
$\beta_j.$ It is assumed that each $\bflam_j$ has dimension greater
than or equal to one, that all blocks have the same dimension, and
that the $\bflam_{j1} = \lambda$ for each $j \geq 1$. Evidently, the
case where dim$(\bflam_j) = 1$ for $j \geq 1$ simply corresponds to
the setting in which each coefficient is penalized in exactly the same
way; permitting the dimension of $\bflam_{j}$ to exceed one allows the
penalty to depend on additional parameters (e.g., weights, such as in
the case of the adaptive lasso considered in \citet{zou:alas}). We are
interested in problems with penalization; therefore, $\lambda$ is
assumed bounded and strictly positive throughout this paper. Several
specific examples will be discussed below.  For any bounded $\bftheta$
with $\lambda>0$ as the first element, and the remainder of $\bftheta$
collecting any additional parameters used to define the penalty, the
scalar function $\tilde{p}(r;\bftheta)$ is assumed to satisfy the
following condition:
\begin{itemize}
\item[(P1)] $\tilde{p}(r;\bftheta) > 0$ for $r > 0$;
  $\tilde{p}(0;\bftheta) = 0;$ $\tilde{p}(r;\bftheta)$ is a
  continuously differentiable concave function with
  $\tilde{p}'(r;\bftheta) \geq 0$ for $r > 0$, and,
  $\tilde{p}'(0+;\bftheta) \in [M^{-1}_{\bftheta},M_{\bftheta}]$ for
  some finite $M_{\bftheta} > 0$.
\end{itemize}
Evidently, (P1) implies that $\tilde{p}'(r;\bftheta) > 0$ for $r\in
(0,K_{\bftheta}),$ where $K_{\bftheta} > 0$ may be finite or
infinite. The combination of the concavity and nonnegative derivative
conditions thus imply that the penalty increases away from the origin,
but with a decreasing rate of growth that may become zero.  The case
where \eqref{eq:gen_pen} is identically zero for $r > 0$ is ruled out
by the positivity of the right derivative at the origin imposed in
(P1); similarly, the concavity assumption also rules out the
possibility of a strictly convex penalty term. Neither of these
restrictions is particularly problematic. Our specific interest lies
in the development of algorithms for estimation problems
subject to a penalty singular at the origin. Were
\eqref{eq:gen_pen} absent, or replaced by a strictly convex penalty
term, the convexity of $g(\bfbeta)$ implies \eqref{the objective} can
be minimized directly using any suitable convex optimization
algorithm, such as that discussed in Theorem
\ref{thm: gen_MIST} below.

Theorem \ref{thm:gen_penal_MM} establishes local convergence of the
indicated class of MM algorithms for minimizing objective functions of
the form \eqref{the objective}.  A proof is provided in Appendix
\ref{app:genpenMM}, where it is shown that conditions imposed in the
statement of the theorem are sufficient conditions for the application
of the general MM local convergence theory summarized in Appendix
\ref{app:MMlocal}.

\begin{thm}
\label{thm:gen_penal_MM}
Let $g(\bfbeta)$ be convex and coercive and assume
$p(\bfbeta;\bflam)$ satisfies both \eqref{eq:gen_pen} and condition
(P1).  Let $h(\bfbeta,\bfalpha) \geq 0$ be a real-valued,
continuous function of $\bfbeta$ and $\bfalpha$ that is
continuously differentiable in $\bfbeta$ for each $\bfalpha$
and satisfies $h(\bfbeta,\bfalpha) = 0$ when $\bfbeta = \bfalpha$.  Let
\benum
\label{q fun}
q(\bfbeta,\bfalpha;\bflam) =
\sum_{j=1}^p \tilde{q}(|\beta_j|,|\alpha_j|; \bflam_j),
\eenum
where $\tilde{q}(r,s; \bftheta) = \tilde{p}(s;\bftheta)+
\tilde{p}'(s; \bftheta)(r-s)$ for $r, s \geq 0$, and define
\[
\psi(\bfbeta,\bfalpha) =
h(\bfbeta,\bfalpha) +
q(\bfbeta,\bfalpha;\bflam) -
p(\bfbeta;\bflam).
\]
Assume the set of stationary points ${\cal S}$ for $\obfn(\bfbeta),
\bfbeta \in {\cal B}$ is finite and isolated. Then:
\bi
\item[(i)] $\obfn(\bfbeta)$ in \eqref{the objective}
is locally Lipschitz continuous and coercive;

\item[(ii)] $q(\bfbeta,\bfalpha;\bflam) - p(\bfbeta;\bflam)$ is either
  identically zero or non-negative for all $\bfbeta \neq \bfalpha;$

\item[(iii)] $\sobfn(\bfbeta,\bfalpha) \equiv \obfn(\bfbeta) +
  \psi(\bfbeta,\bfalpha)$ majorizes $\obfn(\bfbeta)$ and the MM
  algorithm derived from $\sobfn(\bfbeta,\bfalpha) $ converges to a
  stationary point of $\obfn(\bfbeta)$ if $\sobfn(\bfbeta,\bfalpha)$
  is uniquely minimized in $\bfbeta$ for each $\bfalpha$ and at least
  one of $h(\bfbeta,\bfalpha)$ or $q(\bfbeta,\bfalpha;\bflam) -
  p(\bfbeta;\bflam)$ is strictly positive for each $\bfbeta \neq
  \bfalpha.$ \ei
\end{thm}

Condition (iii) of Theorem \ref{thm:gen_penal_MM} establishes
convergence under the assumption that $\sobfn(\bfbeta,\bfalpha)$
strictly majorizes $\obfn(\bfbeta)$ and has a unique minimizer in
$\bfbeta$ for each $\bfalpha$. Such a uniqueness condition is shown by
\cite{vaida} to ensure convergence of the EM and MM algorithms to a
stationary point under more restrictive differentiability
conditions. Importantly, the assumption of globally strict
majorization is only a sufficient condition for convergence; this
condition is only important insofar as it guarantees a strict decrease
in the objective function at every iteration. As can be seen from the
proof, it is possible to relax this condition to locally strict
majorization, in which $\sobfn(\bfbeta,\bfalpha)$ majorizes
$\obfn(\bfbeta),$ with strict majorization being necessary only in an
open neighborhood containing $M(\bfalpha)$.

The use of the MM algorithm and selection of \eqref{q fun} are
motivated by the results \cite{zou:li}; we refer the reader to Remark
\ref{zou li comment} below for further comments in this direction. The
assumptions on $g(\bfbeta)$ clearly cover the case of the linear and
canonically parameterized generalized linear models upon setting
$g(\bfbeta) = - \ell(\bfbeta),$ where $\ell(\bfbeta)$ denotes the
corresponding loglikelihood function. Estimation under the
semiparametric Cox regression model \citep{Cox1972} and accelerated
failure time models are also covered upon setting $g(\bfbeta)$ to be
either the negative logarithm of the partial likelihood function
\citep[e.g., ][Thm VII.2.1]{abgk93} or the Gehan objective function
\citep[e.g., ][]{fygrit94,jstraw09}.

\begin{figure}[!th]
\begin{center}
\includegraphics[angle=270, clip, width=1\textwidth]{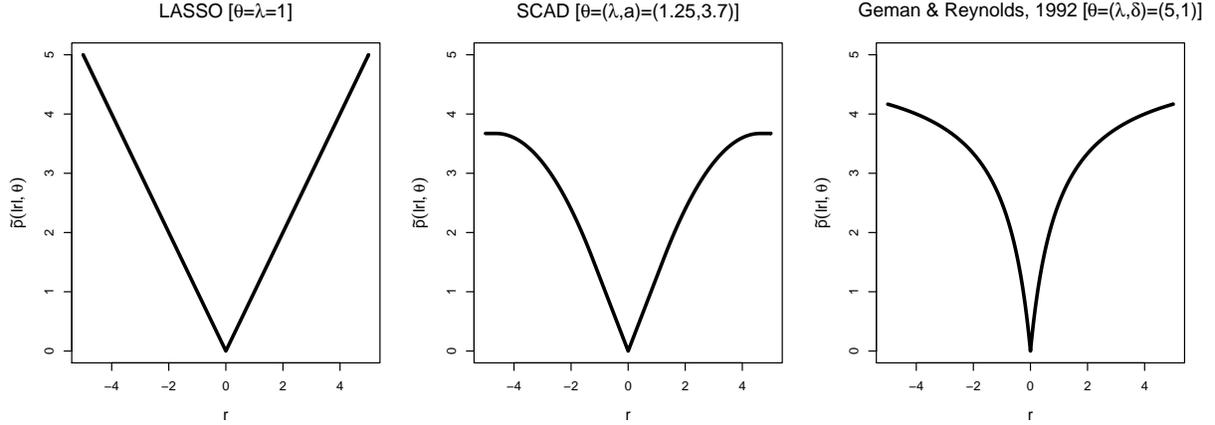}
%\centerline{\includegraphics[angle=270,trim = 5mm 0mm 106mm 3mm, clip, width=1\textwidth]{penalties.epsi}}
\caption{Three examples of penalties satisfying (P1).\label{penalties}}
\end{center}
\end{figure}

The assumption (P1) on the penalty function covers a wide variety of
popular and interesting examples; see Figure{~\ref{penalties}} for
illustration.  For example, the lasso \citep[LAS; e.g.,][]{tibs},
adaptive lasso \citep[ALAS; e.g.,][]{zou:alas}, elastic net \citep[EN;
e.g.,][]{en}, and adaptive elastic net \citep[AEN; e.g.,][]{zou:zhang}
penalties are all recovered as special cases upon considering the
combination of \eqref{eq:gen_pen} and the ridge-type penalty
$\lambda \varepsilon \| \bfbeta \|^2.$ Specifically, with $\bflam_j =
(\lambda,\omega_j)^T$ for $\omega_j \geq 0$, taking
$\tilde{p}(r;\bflam_j) = \lambda \omega_j r$ in \eqref{eq:gen_pen}
gives LAS $(\omega_j = 1, \epsilon = 0),$ ALAS $(\omega_j >0, \epsilon
= 0)$, EN $(\omega_j = 1, \epsilon >0)$ and the AEN $(\omega_j >0,
\epsilon >0 )$ penalties.  It is easy to see that selecting
$\tilde{p}(r;\bflam_j) = \lambda \omega_j r$ also implies the equality
of \eqref{eq:gen_pen} and \eqref{q fun}, a result relevant in both
(ii) and (iii) of Theorem \ref{thm:gen_penal_MM} above.

The proposed penalty specification also covers the smoothly clipped
absolute deviation \citep[SCAD; e.g.,][]{FanLi} penalty upon setting
$\tilde{p}(r;\bflam_j) = \tilde{p}_S(r; \lambda,a)$ for each $j \geq 1$,
where $\tilde{p}_S(r;\lambda,a)$ is defined as the definite
integral of
\begin{equation}
        \tilde{p}'_S(u; \lambda,a) =
\lambda[I(u \leq \lambda) + \frac{(a \lambda - u)_+}{(a-1)\lambda}
I(u>\lambda)]
\end{equation}
on the interval $0 \leq u \leq r$ and some fixed value of $a>2$ (e.g.,
$a=3.7$). The resulting penalty function is continuously
differentiable and concave on $r \in [0,\infty)$.  The concavity of
$\tilde{p}_S(\cdot;\lambda,a)$ on $[0,\infty),$ combined with
$\tilde{p}_S(0;\lambda,a) = 0$ and the fact that
$\tilde{p}'_S(0+;\lambda,a)$ is finite, implies
\benum
\label{concavity}
\tilde{p}_S(r;\lambda,a) \leq \tilde{p}_S(s;\lambda,a) +
\tilde{p}'_S(s;\lambda,a)(r-s)
\eenum
for each $r,s \geq 0$, the boundary cases for $r = 0$ and/or $s = 0$
following from \citet[][Remark 4.1.2, p.\ 21]{hul96}.  In other words,
$\tilde{p}_S(r;\lambda,a)$ can be majorized by a linear function of $r.$

The lasso penalty, its variants, and SCAD have received the greatest
attention in the literature.  More recently, \cite{mcp} introduced the
minimax concave penalty (MCP), which similarly to SCAD is defined in
terms of its derivative.  Specifically, one takes
$\tilde{p}(r;\bflam_j) = \tilde{p}_M(r; \lambda,a)$ for each $j \geq
1$ in \eqref{eq:gen_pen}, where $\tilde{p}_M(r;\lambda,a)$ is defined
for $a>2$ as the definite integral of
\begin{equation}
	\tilde{p}'_{M}(u; \lambda, a) = \left( \lambda - \frac{u}{a} \right)_+
\end{equation}
on the interval $0 \leq u \leq r$ and some fixed value of $a>2$
\citep[e.g., $a=3.7$ as in][]{fan:sam:wu09}.  Further examples of
differentiable concave penalties include $\tilde{p}(r;\bflam_j) =
\tilde{p}_G(r; \lambda, \delta)$ for
\begin{equation}
  \tilde{p}_G(r;\lambda,\delta) = \lambda
\frac{\delta r}{1+\delta r}, ~ \delta > 0
\end{equation}
\citep[e.g.][]{gem:rey, nik}; and $\tilde{p}(r;\bflam_j) =
\tilde{p}_Y(r; \lambda,\delta)$ for
\begin{equation}
\tilde{p}_Y(r;\lambda,\delta) = \lambda \log(\delta r + 1), ~ \delta > 0;
\end{equation}
\citep[e.g.][]{ant}. These penalties represent just a small sample of
the set of possible penalties satisfying (P1) that one might
reasonably consider.

\begin{remark}
  The SCAD and MCP penalties are not strictly concave and
  lead to surrogate majorizers that fail to
  satisfy the globally strict majorization condition in (iii) of
  Theorem \ref{thm:gen_penal_MM} unless $h(\bfbeta,\bfalpha)$ is
  strictly positive whenever $\bfbeta \neq \bfalpha$; see
  Remark \ref{zou li comment} for further discussion and
  also Theorem \ref{prop:sur:glm} below.
\end{remark}

\section{MIST: Minimization by Iterated Soft Thresholding}

\subsection{The MIST algorithm}
In general, the statistical literature on penalized estimation has
proposed optimization algorithms tailored for specific combinations
of fidelity and penalty functions. The class of MM algorithms
suggested by Theorem \ref{thm:gen_penal_MM} provides a very general
and useful framework for proposing new algorithms, the key to
which is a methodology for solving the minimization problem
\eqref{the map}, a step repeated with each iteration of the MM
algorithm.  In this regard, it is helpful to note that the
problem of minimizing $\sobfn(\bfbeta,\bfalpha)$ for a given
$\bfalpha$ is equivalent to minimizing \benum
\label{eq: MIST object 1}
g(\bfbeta) + \lambda \varepsilon \| \bfbeta \|^2_2 +
h(\bfbeta,\bfalpha) + \sum_{j=1}^p \tilde{p}'(|\alpha_j|; \bflam_j)
|\beta_j|
\eenum
in $\bfbeta$. In particular, if $g(\bfbeta) + \lambda \varepsilon \|
\bfbeta \|^2 + h(\bfbeta,\bfalpha)$ is strictly convex for each
bounded $\bfalpha$, which clearly occurs if both $g(\bfbeta)$ and
$h(\bfbeta,\bfalpha)$ are convex in $\bfbeta$ and at least one is
strictly convex, then \eqref{eq: MIST object 1} is also strictly
convex and the corresponding minimization problem has a unique
solution.

\begin{remark}
\label{zou li comment}
For $\varepsilon = h(\bfbeta,\bfalpha) = 0$ and $g(\bfbeta) = -
\ell(\bfbeta)$ for $\ell(\bfbeta) = \sum_{i=1}^n \ell_i(\bfbeta)$ with
$\ell_i(\bfbeta)$ a twice continuously differentiable loglikelihood
function, the majorizer used by the MM algorithm induced by the
surrogate function \eqref{eq: MIST object 1} corresponds (up to sign)
to the minorizer employed in the LLA algorithm of \cite{zou:li}, an
improvement on the so-called LQA algorithm proposed in
\cite{hunter:li}.  \citet[][Proposition 1]{zou:li} assert convergence
of their LLA algorithm under imprecisely stated assumptions and are
additionally unclear as to the nature of convergence result actually
estabished. For example, while \citet[][Theorem 1]{zou:li} demonstrate
that the LLA algorithm does indeed have an ascent property, their
result appears to be insufficient to ensure that the proper analog of
condition Z3(ii) of Theorem \ref{thm: Zangwill} holds in the case of
the SCAD penalty. As a consequence, it is unclear whether even weak
``subsequence'' convergence results \citep[cf.][]{Wu} hold with useful
generality in the case of the LLA algorithm.  In contrast, Theorem
\ref{thm:gen_penal_MM} shows that strict majorization, under a few
precisely stated conditions, is sufficient to ensure local convergence
of the resulting MM algorithm to a stationary point of \eqref{the
  objective} In Section \ref{sec: pen GLM}, it is further demonstrated
how a particular choice of $h(\bfbeta,\bfalpha)$ yields a strict
majorizer that permits both closed form minimization and componentwise
updating at each step of the MM algorithm, even in the case of
penalties that fail to be strictly concave.
\end{remark}

Numerous methods exist for minimizing a differentiable convex
objective function \citep[e.g.,][]{bv04}.  However, because \eqref{eq:
  MIST object 1} is not differentiable, such methods do not apply in
the current setting.  Specialized methods exist for nonsmooth problems
of the form \eqref{eq: MIST object 1} in settings where $g(\bfbeta)$
has a special structure; a well-known example here is LARS
\citep{LARS}, which can be used to efficiently solve lasso-type
problems in the case where $g(\bfbeta)$ is replaced by a least squares
objective function.  Recently, \citet[][Proposition 3.1; Theorem
3.4]{cw05} proposed a very general class of fixed point algorithms for
minimizing $f_1(h) + f_2(h),$ where $f_i(\cdot), i = 1,2$ are each
convex and $h$ takes values in some real Hilbert space ${\cal H}$.
\citet[][Theorem 4.5]{hyz08} specialize the results of \cite{cw05} to
the case where ${\cal H}$ is some subset of $\Real^p$ and $f_2(h) =
\sum_{j=1}^p |h_i|$.  The collective application of these results to
the problem of minimizing \eqref{eq: MIST object 1} generates an
iterated soft-thresholding procedure with an appealingly simple
structure. Theorem \ref{thm: gen_MIST}, given below, states the
algorithm, along with conditions under which the algorithm is
guaranteed to converge; a proof is provided in Appendix
\ref{app:genMIST}. The resulting class of procedures, that is, MM
algorithms in which the minimization of \eqref{eq: MIST object 1} is
carried out via this iterated soft-thresholding procedure, is
hereafter referred to as MIST, an acronym for (M)inimization by
(I)terated (S)oft (T)hresholding. Two important and useful features of
MIST include the absence of high-dimensional matrix inversion and the
ability to update each individual parameter separately.

\begin{thm}
\label{thm: gen_MIST}
Suppose the conditions of Theorem \ref{thm:gen_penal_MM} hold.
Let $m(\bfbeta) = g(\bfbeta) + h(\bfbeta,\bfalpha) + \lambda \epsilon
\| \bfbeta \|^2$ be strictly convex with a Lipschitz continuous
derivative of order $L^{-1} > 0$ for each bounded $\bfalpha.$ Then,
for any such $\bfalpha$ and a constant $\varpi \in (0, 2L)$, the unique
minimizer of \eqref{eq: MIST object 1} can be obtained
in a finite number of iterations using iterated
soft-thresholding:
\begin{enumerate}
\item Set $n = 1$ and initialize $\mathbf{b}^{(0)}$
\item Compute $\mathbf{d}^{(n)} = \mathbf{b}^{(n-1)} -
\varpi \nabla m(\mathbf{b}^{(n-1)})$
\item Compute $\mathbf{b}^{(n)} = S(\mathbf{d}^{(n)};\varpi \bftau)$, where
for any vectors $\mathbf{u},\mathbf{v} \in \Real^p,$
\benum
\label{eq: S fun}
S(\mathbf{u};\mathbf{v}) =
\sum_{j=1}^p s(u_j,v_j) \, \mathbf{e}_j,
\eenum
$\textbf e_j$ denotes the $j^{th}$ unit vector for $\Real^p,$
\begin{equation}\label{soft}
s(u_j,v_j)=\textrm{sign}(u_j)(|u_j| - v_j)_+,
\end{equation}
is the univariate soft-thresholding operator,
and
\[
\bftau =
(\tilde{p}'(|\alpha_1|; \bflam_1),\ldots,\tilde{p}'(|\alpha_p|; \bflam_p))^T.
\]
\item Stop if converged; else, set $n=n+1$ and return to Step 2.
\end{enumerate}
\end{thm}

The proposed algorithm, as originally developed in \cite{cw05}, is
suitable for minimizing the sum of a differentiable convex function
$m(\cdot)$ and another convex function; hence, under similar
conditions, one could employ this algorithm directly to minimize
\eqref{the objective} in cases where the penalty \eqref{eq:gen_pen} is
derived from some convex function $\tilde{p}(\cdot;\bftheta)$.
Theorem 3.4 of \cite{cw05} further shows that the update in Step 3 can
be generalized to
\[
\mathbf{b}^{(n)} = \mathbf{b}^{(n-1)} + \delta_n
\left[
S\left( \mathbf{b}^{(n-1)} - \varpi_n \nabla m(\mathbf{b}^{(n-1)}); \varpi_n \bftau \right)
- \mathbf{b}^{(n-1)} \right],
\]
where, for every $n$, $\varpi_n \in (0,2L)$ and $\delta_n \in (0,1]$
is a suitable sequence of relaxation constants. Judicious selection
of these constants, possibly updated at each step, may improve the
convergence rate of this algorithm.

Theorem \ref{thm: gen_MIST} imposes the relatively strong condition
that the gradient of $m(\bfbeta)$ is $L^{-1}$-Lipschitz continuous.
The role of this condition, also imposed in \citet[][Proposition 3.1;
Theorem 3.4]{cw05}, is to ensure that the update at each step of the
proposed algorithm is a contraction, thereby guaranteeing its
convergence to a fixed point. To see this, note that the update from
$\mathbf{b}^{(n)}$ to $\mathbf{b}^{(n+1)}$ in the algorithm of Theorem
\ref{thm: gen_MIST} involves the mapping $S\!\left( \mathbf{b} -
  \varpi \nabla m(\mathbf{b}); \varpi \bftau \right).$ For any bounded
$\mathbf{b}$ and $\mathbf{a}$, it is easily shown that
\[
\| S\!\left( \mathbf{b} - \varpi \nabla m(\mathbf{b}); \varpi \bftau \right)
- S\!\left( \mathbf{a} - \varpi \nabla m(\mathbf{a}); \varpi \bftau \right)
\|
\leq \|
\mathbf{b} - \mathbf{a} - \varpi \left( \nabla m(\mathbf{b}) -
\nabla m(\mathbf{a}) \right) \|.
\]
When $\nabla m(\mathbf{b}) = \nabla m(\mathbf{a})$, the right-hand
side reduces to $\|\mathbf{b} -\mathbf{a}\|$, and the resulting
mapping is only nonexpansive (i.e., not necessarily
contractive). However, under strict convexity, this situation can
occur only if $\mathbf{b} = \mathbf{a}.$ In particular, suppose that
$\mathbf{b}^{(n)} \neq \mathbf{b}^{(n-1)};$ then, $\nabla
m(\mathbf{b}^{(n)}) \neq \nabla m(\mathbf{b}^{(n-1)})$ and, using the
mean value theorem,
\begin{eqnarray*}
\|\mathbf{b}^{(n+1)} - \mathbf{b}^{(n)} \| & = &
\| S\!\left(\mathbf{b}^{(n)} - \varpi \nabla m(\mathbf{b}^{(n)});
\varpi \bftau \right)
- S\!\left( \mathbf{b}^{(n-1)} - \varpi \nabla m( \mathbf{b}^{(n-1)});
\varpi \bftau \right)
\| \\
& \leq & \| I - \varpi H( \mathbf{b}^{(n)}, \mathbf{b}^{(n-1)}) \| \,
\|  \mathbf{b}^{(n)} -  \mathbf{b}^{(n-1)} \|,
\end{eqnarray*}
where $H(\mathbf{b},\mathbf{a}) = \int_0^1 \nabla m(\mathbf{b} +
t(\mathbf{a}-\mathbf{b})) dt$. The assumption that the gradient of
$m(\bfbeta)$ is $L^{-1}$-Lipschitz continuous now implies that
choosing $\varpi$ as indicated guarantees $\| I - \varpi H(
\mathbf{b}^{(n)}, \mathbf{b}^{(n-1)}) \| < 1$, thereby producing a
contraction.

In view of the generality of the Contraction Mapping Theorem
\citep[e.g., ][Thm.\ 10.2.1]{Luenberger08}, it is possible to relax
the requirement that $\nabla m(\bfbeta)$ is globally
$L^{-1}$-Lipschitz continuous provided that one selects a suitable
starting point.  The relevant extension is summarized in the corollary
below; one may prove this result in a manner similar to Theorem 4.5 of
\cite{hyz08}.

\begin{cor}
\label{cor: gen_MIST}
Let the conditions of Theorem \ref{thm:gen_penal_MM} hold.  Suppose
$\bfalpha$ is a bounded vector and assume that $m(\bfbeta) =
g(\bfbeta) + h(\bfbeta,\bfalpha) + \lambda \epsilon \| \bfbeta \|^2$
is strictly convex and twice continuously differentiable. Then, for a
given bounded $\bfalpha$, there exists a unique minimizer $\bfbeta^*$.
Let $\Omega$ be a bounded convex set containing $\bfbeta^*$ and define
$\lambda_{max}(\bfbeta)$ to be the largest eigenvalue of $\nabla^2
m(\bfbeta)$. Then, the algorithm of Theorem \ref{thm: gen_MIST}
converges to $\bfbeta^*$ in a finite number of iterations provided
that $\mathbf{b}^{(0)} \in \Omega$, $\lambda^* = \max_{\bfbeta \in
  \Omega} \lambda_{max}(\bfbeta) < \infty,$ and $\varpi \in (0,
2/\lambda^*)$.
\end{cor}

Some useful insight into the form of the proposed thresholding
algorithm can be gained by considering the behavior of the penalty
derivative term $\tilde{p}'(r;\bftheta).$ As suggested earlier, (P1)
implies that $\tilde{p}'(r;\bftheta)$ decreases from its maximum value
towards zero as $r$ moves away from the origin. For some penalties
(e.g., SCAD, MCP), this derivative actually becomes zero at some
finite value of $r > 0$, resulting in situations for which $\tau_j =
\tilde{p}'(|\alpha_j|; \bflam_j) = 0$ for at least one $j$.  If
this occurs for component $j$,
then $j^{th}$ component of the vector $S\!\left(\mathbf{b}^{(n)} -
  \varpi \nabla m(\mathbf{b}^{(n)}); \varpi \bftau \right)$ simply
reduces to the $j^{th}$ component of the argument vector
$\mathbf{b}^{(n)} - \varpi \nabla m(\mathbf{b}^{(n)})$.  In the
extreme case where $\bftau = \mathbf{0}$, the proposed update
reduces to $\mathbf{b}^{(n+1)} = \mathbf{b}^{(n)} - \varpi \nabla
m(\mathbf{b}^{(n)}),$ an inexact Newton step in which the inverse
hessian matrix is replaced by $\varpi \bfI_p$, $\bfI_p$ denoting the
$p \times p$ identity matrix, and with step-size chosen to ensure that
this update yields a contraction. Hence, if each of the components of
$\mathbf{b}^{(n)} - \varpi \nabla m(\mathbf{b}^{(n)})$ are
sufficiently large in magnitude, the proposed algorithm simply takes
an inexact Newton step towards the solution; otherwise, one or more
components of this Newton-like update are subject to
soft-thresholding.

\subsection{Penalized estimation for generalized linear models}
\label{sec: pen GLM}

The combination of Theorems \ref{thm:gen_penal_MM}, \ref{thm:
  gen_MIST} and Corollary \ref{cor: gen_MIST} lead to a useful and
stable class of algorithms with the ability to deal with a wide range
of penalized regression problems.  In settings where $g(\bfbeta)$ is
strictly convex and twice continuously differentiable, one can safely
assume that $h(\bfbeta, \bfalpha) = 0$ for all choices of $\bfbeta$
and $\bfalpha$ provided that $\tilde{p}'(r;\bftheta)$ in (P1) is
strictly positive for $r>0$; important examples of statistical
estimation problems here include many commonly used linear and
generalized linear regression models, semiparametric Cox regression
\citep{Cox1972}, and smoothed versions of the accelerated failure time
regression model \citep[cf. ][]{jstraw09}.  The SCAD and MCP
penalizations, as well as other penalties having
$\tilde{p}'(r;\bftheta)\geq 0$ for $r>0,$ can also be used; however,
additional care is required. In particular, and as pointed out in an
earlier remark, if one sets $h(\bfbeta, \bfalpha) = 0$ for all
$\bfbeta$ and $\bfalpha$ then convergence of the resulting algorithm
to a stationary point is no longer guaranteed by the above results due
to the resulting failure of these penalties to induce strict
majorization.

The need to use an iterative algorithm for repeatedly minimizing
\eqref{eq: MIST object 1} is not unusual for the class of MM
algorithms. However, it turns out that for certain choices of
$g(\bfbeta)$, a suitable choice of $h(\bfbeta, \bfalpha)$ in Theorem
\ref{thm: gen_MIST} guarantees both strict majorization and permits
one to minimize \eqref{eq: MIST object 1} in a single iteration,
resulting in a single soft-thresholding update at each
iteration. Below, we demonstrate how the MIST algorithm simplifies in
settings where $g(\bfbeta)$ corresponds to the negative loglikelihood
function of a canonically parameterized generalized linear regression
model having a bounded hessian function. The result applies to all
penalties satisfying condition (P1), including SCAD and MCP.  A proof
is provided in Appendix \ref{app:surglm}.

\begin{thm}\label{prop:sur:glm}
  Let $\bfy$ be $N \times 1$ and suppose the probability distribution
  of $\bfy$ follows a generalized linear model with a canonical link
  and linear predictor $\tbfX \tbfbeta$, where $\tbfX = [ {\textbf
    1}_N, \bfX]$ is $N \times (p+1)$ and $\tbfbeta =
  [\beta_0,\bfbeta^T]^T$ is $(p+1) \times 1$ with $\beta_0$ denoting
  an intercept.  Assume that $g(\tbfbeta) = -\ell(\tbfbeta)$, where
  \[
  \ell(\tbfbeta) =  \textbf{1}^T (c(\bfy) - b(\tbfeta)) + \bfy^T \tbfeta
  \]
  denotes the corresponding loglikelihood; here, $\tbfeta = \tbfX \,
  \tbfbeta$ and $E[y_i] = b'(\tilde{\eta}_i)$ for $i = 1 \ldots N$ for
  $b(\cdot)$ strictly convex and twice continuously differentiable.
  Let the penalty function be defined as in \eqref{eq:gen_pen} and
  satisfy (P1); note that $\beta_0$ remains unpenalized. Define
  \begin{equation}
  \label{h glm}
  h(\tbfbeta,\tbfalpha) =
  \ell(\tbfbeta) - \ell(\tbfalpha) - \nabla\ell(\tbfalpha)^T
  (\tbfbeta-\tbfalpha) +
  \varpi^{-1}(\tbfbeta-\tbfalpha)^T(\tbfbeta-\tbfalpha);
\end{equation}
where $\tilde{\bfalpha}\equiv[\alpha_0,\bfalpha^T]^T$ is $(p+1) \times
1$, and $\varpi$ is defined as in Corollary \ref{cor: gen_MIST}.
Then:
\begin{enumerate}
\item The objective function \eqref{the objective}, say
  $\xi_{glm}(\tilde{\bfbeta}),$ is majorized by
\begin{eqnarray}
\label{eq: MIST_glm}
\nonumber
 \xi^{SUR}_{glm}( \tilde{\bfbeta}, \tilde{\bfalpha}  ) &=&
- \ell(\tilde{\bfalpha}) -\nabla\ell(\tilde{\bfalpha})^T
(\tilde{\bfbeta}-\tilde{\bfalpha}) \\
  && + \varpi^{-1}(\tilde{\bfbeta} - \tilde{\bfalpha})^T(\tilde{\bfbeta}
- \tilde{\bfalpha}) + \sum_{j=1}^{p}(\tau_j|\beta_j| + \gamma_j
+ \lambda\epsilon\beta_j^2)
\end{eqnarray}

where $\tau_j=\tilde{p}'(|\alpha_j|;\bflam_j)$ and $\gamma_j =
\tilde{p}(|\alpha_j|;\bflam_j) -
\tilde{p}'(|\alpha_j|;\bflam_j)|\alpha_j| $ are bounded, nonnegative,
and functionally independent of $\tilde{\bfbeta}.$

\item The functions $g(\tbfbeta) = -\ell(\tilde{\bfbeta})$ and
  $h(\tilde{\bfbeta},\tilde{\bfalpha})$ satisfy the regularity
  conditions of Theorems \ref{thm:gen_penal_MM} and
  \ref{thm: gen_MIST}; hence, the
  corresponding MM algorithm converges to a stationary point of
  \eqref{the objective}.

\item For each bounded $\tilde{\bfalpha}$,
\bi
\item[(a)] the minimizer $\tilde{\bfbeta}^\ast$ of
$\xi^{SUR}_{glm}(\tilde{\bfbeta},
  \tilde{\bfalpha} )$ is unique and satisfies
  \begin{eqnarray}
	\nonumber
	\bfbeta^{\ast} &=& \frac{1}{1+\varpi\lambda\epsilon}
S\left(\bfalpha + \frac{\varpi}{2}[\nabla \ell(\tilde{\bfalpha})]_\mathcal{A}, \frac{\varpi}{2}\bftau \right),\\
	\label{beta.star}
	\beta_0^\ast &=& \alpha_0 + \frac{\varpi}{2}[\nabla \ell(\tbfalpha)]_0
\end{eqnarray}
where $S(\cdot; \cdot)$  is the soft-thresholding operator defined in
\eqref{eq: S fun} and $\mathcal{A}=\{1,\ldots,p\}$.
\item[(b)] for each $\tilde{\bfh}\equiv[\kappa_0,\bfh^T]^T
\in \mathcal{R}^{(p+1)},$
\begin{equation} \label{sur.ineq}
 \xi^{SUR}_{glm}( \tilde{\bfbeta}^\ast + \tilde{\bfh}, \tilde{\bfalpha}  ) \geq  \xi^{SUR}_{glm}( \tilde{\bfbeta}^\ast, \tilde{\bfalpha}  )
  + \varpi^{-1}\left\|\tilde{\bfh}\right\|^2.
\end{equation}
\end{itemize}
\end{enumerate}
\end{thm}

In view of previous results, the result in \# 3 of Theorem
\ref{prop:sur:glm} shows that the resulting MM algorithm takes a very
simple form: given the current iterate $\tbfbeta^{(n)}$,
\begin{enumerate}
\item update the unpenalized intercept $\beta_0^{(n)}:$
\[
\beta_0^{(n+1)}= \beta_0^{(n)} +
\frac{\varpi}{2}\left[ \nabla \ell(\tilde{\bfbeta}^{(n)}) \right]_0
\]
\item update the remaining parameters $\bfbeta^{(n)}$:
\begin{equation}
\label{unified}
\boldsymbol{\beta}^{(n+1)} =
\frac{1}{1+\varpi\lambda\epsilon}
S\left( \boldsymbol{\beta}^{(n)} +
  \frac{\varpi}{2}[\nabla \ell(\tilde{\bfbeta}^{(n)})]_\mathcal{A};
  \frac{\varpi}{2}\bftau^{(n)} \right),
\end{equation}
where $\bftau^{(n)} = (\tilde{p}'(|\beta^{(n)}_1|; \bflam_1),\ldots,
\tilde{p}'(|\beta^{(n)}_p|; \bflam_p))^T$.
\end{enumerate}

The specific choice of function $h(\tbfbeta,\tbfalpha)$ clearly serves
two useful purposes: (i) it leads to componentwise-soft thresholding;
and, (ii) it leads to strict majorization, as is required in condition
(iii) of Theorem{~\ref{thm:gen_penal_MM}}, allowing one to establish
the convergence of MIST for SCAD and other penalties having
$\tilde{p}'(r,\bftheta) = 0$ at some finite $r>0.$

Evidently, the algorithm above immediately covers the setting of
penalized linear regression. For example, suppose that $\bfy$ has been
centered to remove $\beta_0$ from consideration and that the problem
has also been rescaled so that $\bfX$, which is now $N \times p,$
satisfies the indicated conditions. Then, the results of the Theorem
\ref{prop:sur:glm} apply directly with
  \[
  -\ell(\bfbeta) = \frac{1}{2}\left\|\bfX\bfbeta - \bfy \right\|^2,~~
\nabla \ell(\bfbeta)=\bfX^T(\bfy - \bfX\bfbeta),~~
  h(\bfbeta,\bfalpha) =
  \varpi^{-1}
  \| \bfbeta - \bfalpha \|^2 -
  \frac{1}{2}\| \bfX \bfbeta - \bfX \bfalpha \|^2,
  \]
  where $\varpi$ is defined as in Corollary \ref{cor: gen_MIST}.
For the class of adaptive elastic net penalties (i.e.,
$\tilde{p}(r;\bflam_j) = \lambda \omega_j r$ in \eqref{eq:gen_pen}), the
resulting iterative scheme is exactly that proposed in \cite[pg.\ 17]{demol},
specialized to the setting of a Euclidean parameter. In particular,
we have $\tau_j = \lambda \omega_j$ and $\gamma_j=0$ in Theorem
{~\ref{prop:sur:glm}}, and the proposed update reduces to
$$  \boldsymbol{\beta}^{(k+1)} = \frac{1}{\nu+2\lambda\epsilon}
S\left( (\nu\bfI - \textbf{X}'\textbf{X})\ \boldsymbol{\beta}^{(k)} +
  \bfX'\bfy ; \lambda \right),$$ where $\nu = 2\varpi^{-1}.$ Setting
$\nu=1$ and $\epsilon=0$ yields the iterative procedure proposed in
\cite{daub}, provided that $\textbf{X}'\textbf{X}$ is scaled such that
$\bfI -\textbf{X}'\textbf{X}$ is positive definite.  The proposed MIST
algorithm extends these iterative componentwise soft-thresholding
procedures to a much wider class of penalty and data fidelity
functions.

The restriction to canonical generalized linear models in Theorem
\ref{prop:sur:glm} is imposed to ensure strict convexity of the
negative loglikelihood.  Our results are easily modified to handle
non-canonical generalized linear models, provided the negative
loglikelihood remains strictly convex in $\tbfbeta$ and the hessian
can be appropriately bounded. Interestingly, not all canonically
parameterized generalized linear models satisfy the regularity
conditions of Theorem \ref{prop:sur:glm}.  One such important class of
problems is penalized likelihood estimation for Poisson regression
models. For example, in the classical setting of $N$ independent
Poisson observations with $E[Y_i | \tilde{X}_i] = d_i \exp\{
\tilde{x}^T_i\tbfbeta\}$ for a known set of constants $d_1 \ldots
d_N$, we have (i.e., up to irrelevant constants) $\ell(\tbfbeta) =
-\sum_{i=1}^N f_i(\tilde{x}^T_i \tbfbeta),$ where
\[
f_i(u) = d_i e^u - y_i u.
\]
It is easy to see that $\nabla \ell(\tbfbeta),$ hence $\nabla
m(\tbfbeta),$ is locally but not globally Lipschitz continuous; hence,
it is not possible to choose a matrix $\mathbf{C}=\varpi^{-1}\bfI$
such that \eqref{eq: MIST_glm} everywhere majorizes
$\obfn_{glm}(\tbfbeta)$.  Nevertheless, progress remains possible.  For
example, Corollary \ref{cor: gen_MIST} implies that that one can still
use a single update of the form \eqref{unified} provided that a
suitable $\Omega,$ hence $\mathbf{C}$ and $\tbfbeta^{(0)},$ can be
identified.  Alternatively, using results
summarized in \cite{beckeretal}, one can instead majorize
$-\ell(\tbfbeta)$ for any bounded $\bfalpha$ using
\[
k(\tbfbeta,\tbfalpha) = \sum_{j=0}^p k_j(\beta_j;\alpha_j) ~~\mbox{
  for }~~ k_j(\beta_j;\alpha_j) = \sum_{i=1}^n I\{x_{ij} \neq 0\} \,
\theta_{ij} \, f_i\left(\frac{x_{ij}}{\theta_{ij}} (\beta_j -
  \alpha_j) + \tilde{x}^T_i \tbfalpha \right),
\]
where, for every $i$, $\theta_{ij} \geq 0$ are any sequence of
constants satisfying %
$\sum_{j=0}^p \theta_{ij} = 1$ and $\theta_{ij} > 0$ if $x_{ij} \neq
0.$ Of importance here is the fact $k_j(\beta_j;\alpha_j)$ is a
strictly convex function of $\beta_j$ and does not depend on
$\beta_{k}$ for $k \neq j$. One may now take $h(\tbfbeta,\tbfalpha)$
in Theorem \ref{thm:gen_penal_MM} as being equal to
$k(\tbfbeta,\tbfalpha) + \ell(\tbfbeta),$ leading to the minimization
of \benum
\label{eq: MM poisson}
\sobfn(\tbfbeta,\tbfalpha) \propto \sum_{j=1}^p [k_j(\beta_j;\alpha_j)
+ \lambda \varepsilon \beta^2_j + \tilde{p}'(|\alpha_j|; \bflam_j)
|\beta_j|] + k_0(\beta_0,\alpha_0).
\eenum
In particular, componentwise soft-thresholding is replaced by
componentwise minimization of \eqref{eq: MM poisson}, the latter
being possible using any algorithm capable of minimizing a
continuous nonlinear function of one variable.

\begin{remark}
  The Cox proportional hazards model \citep{Cox1972}, while not a
  generalized linear model, shares the essential features of the
  generalized linear model utilized in Theorem \ref{prop:sur:glm}. In
  particular, the negative log partial likelihood, say $g(\bfbeta) =
  -\ell_p(\bfbeta),$ is strictly convex, twice continuously
  differentiable, and has a bounded hessian \citep[e.g.,][]{quadbound,
    abgk93}. Consequently, Theorem{~\ref{prop:sur:glm}} and its proof
  are easily modified for this setting upon taking $g(\bfbeta)$ as
  indicated, setting $h(\bfbeta,\bfalpha)= \ell_p(\bfbeta) -
  \ell_p(\bfalpha) - \nabla\ell_p(\bfalpha)^T (\bfbeta-\bfalpha) +
  \varpi^{-1}\|\bfbeta-\bfalpha\|^2,$ and taking $\varpi$ as defined
  as in Corollary \ref{cor: gen_MIST}.
\end{remark}

\subsection{Accelerating Convergence}
\label{sec: cvg acc}

Similarly to the EM algorithm, the stability and simplicity of the MM algorithm
frequently comes at the price of a slow convergence rate.  Numerous methods of
accelerating the EM algorithm have been proposed in the literature;
see \cite{embook} for a review.  Recently, \cite{var:rol} proposed a new
method for EM called SQUAREM, obtained by ``squaring'' an iterative
Steffensen-type (STEM) acceleration method.  Both STEM and SQUAREM are
structured for use with iterative mappings of the form $\theta_{n+1} =
M(\theta_n),$ $n=0,1,2,\ldots,$ hence applicable to both the EM and MM
algorithms.  Specifically, the acceleration update for SQUAREM is
given by
\begin{eqnarray}
\nonumber
\theta_{n+1} &=& \theta_n -2 \gamma_n(M(\theta_n)-\theta_n) + \gamma_n^2[M(M(\theta_n))-2M(\theta_n) + \theta_n] \\
\label{squarem}
&=& \theta_n - 2\gamma_n r_n + \gamma_n^2 v_n,
\end{eqnarray}
where $r_n = M(\theta_n)-\theta_n$ and $v_n =
(M(M(\theta_n))-M(\theta_n)) - r_n$ for an adaptive steplength
$\gamma_n.$ \cite{var:rol} suggest several steplength options, with
preference for choice $\gamma_n = - \|r_n\| / \|v_n\|.$
\cite{rol:var} provide a proof of local convergence for SQUAREM under
restrictive conditions on the EM mapping $M(\theta)$, while
\cite{var:rol} outline a proof for global convergence for
versions of SQUAREM that employ a back-tracking strategy.
We study the effectiveness of SQUAREM applied to the simplified
form of the MIST algorithm, hereafter denoted SQUAREM$^2$,
in Section \ref{sec: cvg acc sim}.

\section{Simulation Results}

The simulation results summarized below are intended to compare the
estimates of $\bfbeta$ obtained from existing methods to those
obtained using the simplified MIST algorithm of Theorem
\ref{prop:sur:glm}. In particular, we consider the performance of MIST
for the class of penalized linear and generalized linear models,
demonstrating its capability of recovering the solutions
provided by existing algorithms when both algorithms are forced to use
the same set of ``tuning'' parameters (i.e., penalty parameter(s),
plus any additional parameters required to define the penalty itself).
In cases where multiple local minima can arise, we further show that
the MIST algorithm often tends to find solutions with lower objective
function evaluations in comparison with existing algorithms, provided
these algorithms utilize the same choice of starting value.

\subsection{Example 1: Linear Model}

Let $\textbf{1}_m$ and $\textbf{0}_m$ respectively denote
$m$-dimensional vectors of ones and zeros. Then, following
\cite{zou:zhang}, we generated data from the linear regression model
\begin{equation}\label{datagen}
 y = \textbf{x}'\bfbeta^\ast + \varepsilon
\end{equation}
where $\bfbeta^{\ast} = (3\cdot\textbf{1}^T_q, \textbf{0}^T_{p-q})^T$ is a
$p$-dimensional vector with intrinsic dimension $q = 3[p/9],$
$\varepsilon\sim N(0, \sigma^2),$ and $\textbf{x}$ follows a
$p$-dimensional multivariate normal distribution with zero mean and
covariance matrix $\Sigma$ having elements $\Sigma_{j,k} =
\rho^{|j-k|},$ $1 \leq k, j \leq p.$ We considered $\sigma\in\{1,3\},$
$\rho \in \{0.0,0.5, 0.75\}$ and $p \in \{35,81\}$ for $N = 100$
independent observations.

Penalized least squares estimation is considered for five popular
choices of penalty functions, all of which are currently implemented
in the \R software language \citep{R}: LAS, ALAS, EN, AEN, and
SCAD. The LAS, ALAS, EN and AEN penalties are all convex and lead to
unique solutions under mild conditions; the SCAD penalty is concave
and the resulting minimization problem may have multiple solutions. In
each case, we used existing software for computing solutions
subject to these penalizations and compared those results to the
solutions computed using the MIST algorithm.

Regarding existing methods, we respectively used the \textit{lars}
\citep{Rlars} and \textit{elasticnet} \citep{Relasticnet} packages
for computing solutions in the case of the LAS and EN penalties.
For the ALAS and AEN penalties, we used software kindly provided by
\cite{zou:zhang} that also makes use of the \textit{elasticnet} package.
The weights for the AEN penalty are computed using $\omega_j =
|\hat{\beta}^{EN}_j|^{-\gamma},$ $j=1,\ldots,p,$ where
$\hat{\bfbeta}^{EN}$ is an EN estimator and $\gamma$ is a positive
constant.  Using EN-based weights in the AEN fitting algorithm
necessitates tuning parameter specification for both EN and AEN.  As
in \cite{zou:zhang}, the $\ell_1$ parameters $\lambda$ ($\lambda_1$ in
their notation) are allowed to differ, whereas the $\ell_2$ parameters
$\epsilon$ ($\lambda_2$ in their notation) are forced to be the same.
Evidently, setting $\epsilon=0$ ($\lambda_2=0$) results in the ALAS
solution. For the SCAD penalty, we considered the estimator of
\cite{kim08} (HD), as well the one-step SCAD (1S) and LLA estimators
of \cite{zou:li}.  The code for the first two methods was kindly
provided by their respective authors; the LLA estimator was
computed using the \R package \textit{SIS}. The choice
$a=3.7$ was used for all implementations of SCAD.

We considered finding solutions using penalties in the set
$\Lambda=\{0.1,1,5,10,20,100\}.$ In particular, for LAS and SCAD,
$\lambda=\lambda_1\in\Lambda.$ For EN, both
$\lambda=\lambda_1\in\Lambda$ and
$\lambda\epsilon=\lambda_2\in\Lambda.$ For simplicity, we fixed the
weights for AEN for a given $\lambda_2$ by selecting the `best'
$\hat{\bfbeta}^{EN}$ among the six estimators involving
$\lambda=\lambda_1\in\Lambda$ based on a BIC-like criteria.  Likewise
for ALAS, the weights were computing using the `best'
$\hat{\bfbeta}^{LAS}$ among the six estimators involving
$\lambda=\lambda_1\in\Lambda$.  The parameter $\gamma$ for the ALAS
and AEN penalties was respectively set to three
and five for $p = 35$ and $p =81$.

For the strictly convex objective functions associated with the LAS,
ALAS, EN, and AEN penalties, we simply used a starting value of
$\bfbeta^{(0)}=\textbf{0}_p.$ For SCAD, three different starting
values for the MIST, HD, and LLA SCAD algorithms were considered:
$\bfbeta^{(0)}=\textbf{0}_p,$ $\bfbeta^{(0)}=\widehat{\bfbeta}_{ml}$
(i.e., the unpenalized least squares estimate), and
$\bfbeta^{(0)}=\widehat{\bfbeta}_{1S,\lambda}$ (i.e., the one-step
estimate computed using the penalty $\lambda$).  As in \cite{zou:li},
the one-step estimator 1S is computed using $\widehat{\bfbeta}_{ml}$,
an appropriate choice when $N > p$.

The convergence criteria used by the existing software packages were
used without alteration in this simulation study.  The convergence
criteria used for MIST were as follows: the algorithm stopped if
either (i) the normed difference of successive iterates was less than
$10^{-6}$ (convergence of coefficients); or, (ii) the difference of
the objective function evaluated at successive iterates was less than
$10^{-6}$ and the number of iterations exceeded $10^{6}$ (convergence
of optimization).  Due to the large number of comparisons and highly
intensive nature of the computations, we ran $B=100$ simulations for
each choice of $\rho$, $\sigma$, and $p$. We report the results for
the convex penalties in Table{~\ref{ngp:match:mn}} and those for the
SCAD penalty in Tables {~\ref{ngp.p35.sig13}}
and{~\ref{ngp.p81.sig13}}.

In Table{~\ref{ngp:match:mn}}, we summarize the average normed difference
between the solution obtained using existing software and that
obtained using MIST, $ \left\| \hat{\bfbeta}_{exist} -
  \hat{\bfbeta}_{mist}\right\|,$ over the $B=100$ simulations; in particular,
we report in the two leftmost panels the maximum value of this difference,
computed across all combinations of tuning parameters.
These maximum differences (all of which
are multiplied by $10^5$) are remarkably small for all (A)LAS and
(A)EN penalties, indicating that MIST recovers the same (unique)
solutions as the existing algorithms.  Interestingly, the values for
LAS are slightly larger than the rest, where the maximum differences
all resulted from the smallest value of $\lambda$ considered
($\lambda=0.1)$.  In these cases, the algorithm tended to stop using
the objective function criteria rather than the (stricter) coefficient
criteria, resulting in slightly larger differences on average.

\begin{table}[!t]
\begin{center}
\caption{{\small Maximum average normed differences ($\times 10^5$) over $B=100$ simulations for Examples 1 (LM) and 2 (GLM).}}
\begin{tabular}{rrrr|rrr||rrr}
  \hline
 & \multicolumn{3}{c|}{$LM:\sigma=1$} & \multicolumn{3}{c||}{$LM:\sigma=3$} & \multicolumn{3}{c}{$GLM$}\\
  \cline{2-4} \cline{5-7} \cline{8-10}
 $\rho$ & 0 & 0.5 & 0.75 & 0 & 0.5 & 0.75 & 0 & 0.5 & 0.75 \\
  \hline
  $p=35$ &  &  &  &  &  & & $q=25$& & \\
  LAS & 0.10 & 0.35 & 1.45 & 0.10 & 0.37 & 1.56 & 0.07 & 4.28 & 6.17 \\
  ALAS & 0.03 & 0.14 & 0.64 & 0.05 & 0.21 & 1.00 & 1.84 & 2.86 & 3.76 \\
  EN & 0.07 & 0.19 & 0.50 & 0.07 & 0.20 & 0.51 & 2.30 & 5.61 & 8.68\\
  AEN & 0.03 & 0.10 & 0.33 & 0.04 & 0.13 & 0.36  & 1.47 & 3.35 & 5.27 \\
    $p=81$ &  &  &  &  &  & &$q=75$ & &  \\
  LAS & 1.73 & 3.82 & 11.76 & 2.33 & 5.78 & 18.99 & 0.10 & 6.97 & 9.94\\
  ALAS & 0.12 & 0.38 & 1.58 & 0.35 & 1.03 & 4.39 & 1.34 & 2.55 & 3.30 \\
  EN & 0.31 & 0.49 & 0.87 & 0.31 & 0.49 & 0.88 & 2.35 & 4.64 & 6.56 \\
  AEN & 0.14 & 0.22 & 0.56 & 0.16 & 0.26 & 0.56 & 1.27 & 2.29 & 2.85 \\
   \hline
\end{tabular}
\label{ngp:match:mn}
\end{center}
\end{table}

The results for SCAD are reported in Tables{~\ref{ngp.p35.sig13}}
($p=35$) and{~\ref{ngp.p81.sig13}} ($p=81$) and display (i) the
average normed differences, multiplied by $10^3,$ for each combination
of $\lambda$, $\rho$, $\sigma$, $p$ and starting value; and, (ii) the
proportion of simulated datasets in which the MIST solution yields a
lower evaluation of the objective function in comparison with the
solution obtained using another method for the indicated choice of
starting value.  The results for $\lambda=100$ are not shown, as the
solution was $\textbf{0}_p$ in all cases. In comparison with the
convex penalties, larger normed differences are observed, even when
controlling for the use of the same starting value. Such differences
are a result of two important features of the SCAD optimization
problem: (i) the possible existence of several local minima; and, (ii)
the fact that the MIST, HD, and LLA algorithms each take a different
path from a given starting value towards one of these solutions. For
example, while each of the LLA, MIST, and HD algorithms involve
majorization of the objective function using a lasso-type surrogate
objective function, both the majorization and minimization of the
resulting surrogate function are carried out differently in each
case. In particular, the LLA algorithm, as implemented in
\textit{SIS}, majorizes only the penalty term and adapts the lasso
code in \textit{glmpath} in order to minimize the corresponding
surrogate objective function at each step. The HD algorithm is similar
in spirit, but instead decomposes the penalty term into a sum of a
concave and convex function and utilizes the the algorithm of
\cite{Rosset} to minimize the corresponding surrogate objective
function. The MIST algorithm instead uses the same penalty
majorization as the LLA algorithm, but additionally majorizes the
negative loglikelihood term in a way that permits minimization of the
surrogate function in a single soft-thresholding step. Each procedure
therefore takes a different path towards a solution, even when given
the same starting value.

We remark here that differences must also expected between any of LLA,
HD, MIST and the one-step solution 1S; from an optimization
perspective, the one-step estimate is the result of running just one
iteration of the LLA algorithm, starting from the unpenalized least
squares estimator $\widehat{\bfbeta}_{ml}$ \citep{zou:li}, and only
provides an approximation the solution to the desired minimization
problem. All other methods (LLA, MIST, HD) iterate until some local
minimizer (or stationary point) is reached.  For example, when using
either $\widehat{\bfbeta}_{ml}$ or $\widehat{\bfbeta}_{1S,\lambda}$ as
the starting value, MIST always found a solution that produced a lower
evaluation of the objective function in comparison to
$\widehat{\bfbeta}_{1S,\lambda}$. However, when using the null
starting value of $\textbf{0}_p$, the one-step estimator did
occasionally result in a lower objective function evaluation in cases
involving smaller values of $\lambda$. This behavior is not terribly
surprising; with small $\lambda,$ the one-step solution should
generally be close to the unpenalized least squares solution, as the
objective function itself is likely to be dominated by the least
squares term.

Of all the SCAD algorithms considered here, MIST and LLA tended to
find the most similar solutions (i.e., have the smallest normed
differences). For the cases in which the LLA solution had lower
objective function evaluations, all of the MIST solutions were also
LLA solutions; i.e, when starting the LLA algorithm with the MIST
solution, the algorithm terminated at the starting value (i.e., the
LLA solution coincides with the MIST solution).  With the exception of
three of these cases, starting the MIST algorithm with the LLA
solution also resulted in the same behavior.  For the most part, the
HD and MIST algorithms also gave similar results, with one source of
difference being the respective stopping criteria used. The stopping
criteria for HD, assessed in order, are as follows: (1) `convergence
of optimization': stop if the absolute value of the difference of the
objective evaluated at successive iterates is less than 1e-6; (2)
`convergence of penalty gradient': stop if the sum of the absolute
value of the differences of the derivative of the centered penalty
evaluated at successive iterates is less than 1e-6, (3) `convergence
of coefficients:' stop if the sum of the absolute value of the
differences of successive iterates is less than 1e-6, and (4)
`jump-over' criteria: stop if the objective at the previous iterate
plus 1e-6 was less than the objective at the current iterate.  After
careful analysis of the results, we can assert the following:
\begin{itemize}
\item The MIST solution usually has the same or a lower evaluation of
  the objective function in comparison with HD, regardless of starting value.

\item HD tends to have the greatest difficulty in cases of high
  correlation between predictors, a likely result of the fact that
  this algorithm relies on the variance of the unpenalized least
  squares estimator, hence matrix inversion, to take steps towards
  solution. In contrast, MIST requires no matrix inversion.
\end{itemize}

On balance, the MIST algorithm performs as well or better than LLA and
HD in locating minimizers in nearly all cases.  As suggested above, variation
in the solutions found can be traced to the path each algorithm takes towards
a solution and differences in stopping criteria.
Remarkably, in cases when the correlation among predictors was
low, the choice of starting value made little difference for MIST; either
the same solution was found for all starting values or none of the starting
values dominated in terms of finding the lower or equivalent objective
evaluations.  In settings involving higher
correlation, however, using either $\textbf{0}_p$ or the $1S$ starting
values tended to result in the lower evaluations of the objective
function in comparison with using the unpenalized least squares
solution.  Similar behavior was observed for the LLA algorithm. In
contrast, the choice of starting value had a much larger impact on the
performance of the HD estimator; in particular, the use of
$\textbf{0}_p$ as a starting value typically resulted in the lowest objective
function evaluations when compared to using
a non-null starting value.

\begin{table}[!ht]
\begin{center}
\caption{{\small Algorithm performance in Example 1 (LM: $p=35, N=100$) for SCAD penalty. The column `avg' is the average normed differences $\times 10^3$ between the MIST solution and the existing method's solution ; `prop' is the proportion of MIST solutions whose objective function evaluation was less than or equal to that of the existing method's solution.}}
{\scriptsize
\begin{tabular}{@{}llrr|rr|rr||rr|rr|rr@{}}
  \hline  \hline
        & & \multicolumn{6}{c||}{$\sigma=1$} & \multicolumn{6}{c}{$\sigma=3$} \\ \cline{3-14}
	$\bfbeta^{(0)}$  & & \multicolumn{2}{c|}{$\textbf{0}_p$} & \multicolumn{2}{c|}{$\widehat{\bfbeta}_{ml}$}
	& \multicolumn{2}{c||}{$\widehat{\bfbeta}_{1S,\lambda}$} & \multicolumn{2}{c|}{$\textbf{0}_p$} & \multicolumn{2}{c|}{$\widehat{\bfbeta}_{ml}$}
	& \multicolumn{2}{c}{$\widehat{\bfbeta}_{1S,\lambda}$}\\
	\cline{3-14}
\multicolumn{2}{l}{$\rho~~$ method} & avg & prop & avg & prop & avg & prop & avg & prop & avg & prop & avg & prop \\
  \hline
  \multicolumn{2}{l}{$\lambda=.1$} &  &  &  & &  &  &  & &  &  &  & \\
  0 & HD & 15.71 & 1.00 & 15.41 & 1.00 & 17.93 & 1.00 & 468.55 & 1.00 & 2076.40 & 1.00 & 55.17 & 1.00 \\
  & 1S & 99.13 & 1.00 & 99.13 & 1.00 & 99.13 & 1.00 & 211.17 & 1.00 & 211.16 & 1.00 & 211.16 & 1.00 \\
  & LLA & 0.43 & 1.00 & 0.46 & 1.00 & 0.46 & 1.00 & 2.07 & 1.00 & 1.96 & 1.00 & 2.02 & 1.00 \\
  0.5 & HD & 7.07 & 0.99 & 10.72 & 1.00 & 2.04 & 1.00 & 269.85 & 0.97 & 218.94 & 0.94 & 130.76 & 0.98 \\
  & 1S & 192.22 & 1.00 & 192.01 & 1.00 & 192.00 & 1.00 & 483.89 & 0.98 & 421.17 & 1.00 & 419.15 & 1.00 \\
  & LLA & 6.65 & 0.99 & 0.62 & 1.00 & 0.60 & 1.00 & 57.87 & 0.96 & 12.84 & 0.99 & 2.37 & 1.00 \\
  0.75 & HD & 29.25 & 0.99 & 105.39 & 0.92 & 66.83 & 0.96 & 2335.23 & 1.00 & 2758.43 & 0.98 & 2731.10 & 0.99 \\
  & 1S & 575.09 & 1.00 & 488.09 & 1.00 & 486.19 & 1.00 & 1417.97 & 0.86 & 604.26 & 1.00 & 629.21 & 1.00 \\
  & LLA & 23.81 & 0.98 & 23.34 & 0.99 & 1.67 & 0.99 & 558.56 & 0.73 & 69.30 & 0.96 & 44.87 & 0.98 \\
 \hline
  \multicolumn{2}{l}{$\lambda=1$} & &  &  &  &  &  &  &  & \\
  0 & HD & 6.22 & 1.00 & 22.87 & 1.00 & 19.99 & 1.00 & 9.44 & 1.00 & 35.16 & 1.00 & 14.65 & 1.00 \\
  & 1S & 694.59 & 1.00 & 694.57 & 1.00 & 694.57 & 1.00 & 844.68 & 1.00 & 844.67 & 1.00 & 844.67 & 1.00 \\
  & LLA & 1.64 & 1.00 & 1.71 & 1.00 & 1.74 & 1.00 & 1.47 & 1.00 & 1.47 & 1.00 & 1.43 & 1.00 \\
  0.5 & HD & 300.62 & 0.98 & 34.09 & 1.00 & 115.76 & 0.98 & 303.98 & 0.96 & 140.26 & 1.00 & 94.90 & 1.00 \\
  & 1S & 4489.01 & 1.00 & 4276.77 & 1.00 & 4261.64 & 1.00 & 3547.69 & 1.00 & 3254.16 & 1.00 & 3254.16 & 1.00 \\
  & LLA & 296.53 & 0.98 & 7.10 & 1.00 & 88.14 & 0.98 & 248.82 & 0.96 & 2.66 & 1.00 & 2.66 & 1.00 \\
  0.75 & HD & 3083.00 & 0.68 & 1980.40 & 0.89 & 1138.53 & 0.96 & 1476.59 & 0.84 & 1669.60 & 0.93 & 868.21 & 0.97 \\
  & 1S & 7224.77 & 1.00 & 5491.09 & 1.00 & 5622.21 & 1.00 & 5682.04 & 0.96 & 3835.30 & 1.00 & 3748.35 & 1.00 \\
  & LLA & 2802.66 & 0.66 & 1121.80 & 0.85 & 293.50 & 0.96 & 1365.76 & 0.83 & 918.63 & 0.89 & 433.66 & 0.96 \\
   \hline
\multicolumn{2}{l}{$\lambda=5$} & &  &  &  &  &  &  &  & \\
  0 & HD & 18.18 & 1.00 & 18.18 & 1.00 & 18.18 & 1.00 & 17.73 & 1.00 & 17.73 & 1.00 & 17.73 & 1.00 \\
  & 1S & 48.23 & 1.00 & 48.23 & 1.00 & 48.23 & 1.00 & 63.63 & 1.00 & 63.63 & 1.00 & 63.63 & 1.00 \\
  & LLA & 0.00 & 1.00 & 0.00 & 1.00 & 0.00 & 1.00 & 0.00 & 1.00 & 0.00 & 1.00 & 0.00 & 1.00 \\
  0.5 & HD & 0.01 & 1.00 & 0.01 & 1.00 & 0.01 & 1.00 & 0.01 & 1.00 & 0.01 & 1.00 & 0.01 & 1.00 \\
  & 1S & 3696.85 & 1.00 & 3696.85 & 1.00 & 3696.85 & 1.00 & 3751.96 & 1.00 & 3751.96 & 1.00 & 3751.96 & 1.00 \\
  & LLA & 0.02 & 1.00 & 0.09 & 1.00 & 0.08 & 1.00 & 0.03 & 1.00 & 0.14 & 1.00 & 0.08 & 1.00 \\
  0.75 & HD & 0.27 & 1.00 & 0.27 & 1.00 & 98.05 & 1.00 & 19.20 & 0.99 & 19.21 & 0.99 & 99.95 & 0.99 \\
  & 1S & 3977.93 & 1.00 & 3977.93 & 1.00 & 4045.81 & 1.00 & 4170.49 & 1.00 & 4170.49 & 1.00 & 4180.79 & 1.00 \\
  & LLA & 0.27 & 1.00 & 0.45 & 1.00 & 98.35 & 1.00 & 19.00 & 0.99 & 19.20 & 0.99 & 100.05 & 0.99 \\
   \hline
\multicolumn{2}{l}{$\lambda=10$} & &  &  &  &  &  &  &  & \\
  0 & HD & 0.00 & 1.00 & 0.00 & 1.00 & 0.00 & 1.00 & 0.00 & 1.00 & 0.00 & 1.00 & 0.00 & 1.00 \\
  & 1S & 0.00 & 1.00 & 0.00 & 1.00 & 0.00 & 1.00 & 0.00 & 1.00 & 0.00 & 1.00 & 0.00 & 1.00 \\
  & LLA & 0.00 & 1.00 & 0.00 & 1.00 & 0.00 & 1.00 & 0.00 & 1.00 & 0.00 & 1.00 & 0.00 & 1.00 \\
  0.5 & HD & 57.33 & 1.00 & 57.33 & 1.00 & 57.33 & 1.00 & 53.80 & 1.00 & 53.80 & 1.00 & 53.80 & 1.00 \\
  & 1S & 501.86 & 1.00 & 501.86 & 1.00 & 501.86 & 1.00 & 497.87 & 1.00 & 497.87 & 1.00 & 497.87 & 1.00 \\
  & LLA & 0.01 & 1.00 & 0.03 & 1.00 & 0.01 & 1.00 & 0.01 & 1.00 & 0.04 & 1.00 & 0.01 & 1.00 \\
  0.75 & HD & 0.41 & 1.00 & 0.41 & 1.00 & 0.41 & 1.00 & 0.53 & 1.00 & 0.53 & 1.00 & 0.53 & 1.00 \\
  & 1S & 4206.65 & 1.00 & 4206.65 & 1.00 & 4206.65 & 1.00 & 4261.12 & 1.00 & 4261.12 & 1.00 & 4261.12 & 1.00 \\
  & LLA & 0.09 & 1.00 & 0.30 & 1.00 & 0.14 & 1.00 & 0.07 & 1.00 & 0.36 & 1.00 & 0.10 & 1.00 \\
  \hline
 \multicolumn{2}{l}{$\lambda=20$} & &  &  &  &  &  &  &  & \\
  0 & HD & 0.00 & 1.00 & 0.00 & 1.00 & 0.00 & 1.00 & 0.00 & 1.00 & 0.00 & 1.00 & 0.00 & 1.00 \\
  & 1S & 0.00 & 1.00 & 0.00 & 1.00 & 0.00 & 1.00 & 0.00 & 1.00 & 0.00 & 1.00 & 0.00 & 1.00 \\
  & LLA & 0.00 & 1.00 & 0.00 & 1.00 & 0.00 & 1.00 & 0.00 & 1.00 & 0.00 & 1.00 & 0.00 & 1.00 \\
  0.5 & HD & 0.00 & 1.00 & 0.00 & 1.00 & 0.00 & 1.00 & 0.00 & 1.00 & 0.00 & 1.00 & 0.00 & 1.00 \\
  & 1S & 0.00 & 1.00 & 0.00 & 1.00 & 0.00 & 1.00 & 0.00 & 1.00 & 0.00 & 1.00 & 0.00 & 1.00 \\
  & LLA & 0.00 & 1.00 & 0.00 & 1.00 & 0.00 & 1.00 & 0.00 & 1.00 & 0.00 & 1.00 & 0.00 & 1.00 \\
  0.75 & HD & 33.90 & 1.00 & 33.90 & 1.00 & 33.90 & 1.00 & 35.46 & 1.00 & 35.46 & 1.00 & 35.46 & 1.00 \\
  & 1S & 47.21 & 1.00 & 47.21 & 1.00 & 47.21 & 1.00 & 46.90 & 1.00 & 46.90 & 1.00 & 46.90 & 1.00 \\
  & LLA & 0.00 & 1.00 & 0.00 & 1.00 & 0.00 & 1.00 & 0.00 & 1.00 & 0.06 & 1.00 & 0.00 & 1.00 \\
\hline  \hline
 \end{tabular}}
\label{ngp.p35.sig13}
\end{center}
\end{table}

\begin{table}[!ht]
\begin{center}
\caption{{\small Algorithm performance in Example 1 (LM: $p=81, N=100$) for SCAD penalty. The column `avg' is the average normed differences ($\times 10^3$) between the MIST solution and the existing method's solution ; `prop' is the proportion of MIST solutions whose objective function evaluation was less than or equal to that of the existing method's solution.}}
{\scriptsize
\begin{tabular}{@{}l@{}lrr|rr|rr||rr|rr|rr@{}}
  \hline  \hline
        & & \multicolumn{6}{c||}{$\sigma=1$} & \multicolumn{6}{c}{$\sigma=3$} \\ \cline{3-14}
	$\bfbeta^{(0)}$  & & \multicolumn{2}{c|}{$\textbf{0}_p$} & \multicolumn{2}{c|}{$\widehat{\bfbeta}_{ml}$}
	& \multicolumn{2}{c||}{$\widehat{\bfbeta}_{1S,\lambda}$} & \multicolumn{2}{c|}{$\textbf{0}_p$} & \multicolumn{2}{c|}{$\widehat{\bfbeta}_{ml}$}
	& \multicolumn{2}{c}{$\widehat{\bfbeta}_{1S,\lambda}$}\\
	\cline{3-14}
\multicolumn{2}{l}{$\rho$ method} & avg & prop & avg & prop & avg & prop & avg & prop & avg & prop & avg & prop \\
  \hline
  \multicolumn{2}{l}{$\lambda=.1$} &  &  &  & &  &  &  & &  &  &  & \\
  0 & HD & 828.22 & 1.00 & 1211.97 & 1.00 & 962.10 & 1.00 & 4615.10 & 1.00 & 5414.49 & 1.00 & 5350.54 & 1.00 \\
  & 1S & 753.85 & 1.00 & 753.84 & 1.00 & 753.84 & 1.00 & 2836.29 & 0.90 & 1314.46 & 1.00 & 1366.62 & 1.00 \\
  & LLA & 1.60 & 1.00 & 1.67 & 1.00 & 1.64 & 1.00 & 1181.62 & 0.76 & 382.17 & 0.82 & 223.32 & 0.94 \\
  0.5 & HD & 5992.88 & 1.00 & 6008.14 & 1.00 & 5994.86 & 1.00 & 8002.08 & 1.00 & 9530.30 & 1.00 & 9546.21 & 1.00 \\
  & 1S & 1217.02 & 1.00 & 1202.01 & 1.00 & 1201.30 & 1.00 & 4619.22 & 0.88 & 1473.61 & 1.00 & 1403.36 & 1.00 \\
  & LLA & 24.78 & 0.97 & 1.33 & 1.00 & 8.50 & 0.99 & 2123.22 & 0.57 & 576.65 & 0.83 & 232.10 & 0.91 \\
  0.75$~$ & HD & 12018.61 & 1.00 & 12042.97 & 1.00 & 12042.90 & 1.00 & 13582.93 & 1.00 & 16580.85 & 1.00 & 16569.80 & 1.00 \\
  & 1S & 2492.18 & 1.00 & 2327.76 & 1.00 & 2330.54 & 1.00 & 8204.45 & 0.60 & 1215.98 & 1.00 & 1181.16 & 1.00 \\
  & LLA & 36.95 & 0.98 & 90.89 & 0.97 & 90.69 & 0.96 & 3517.93 & 0.50 & 607.08 & 0.78 & 252.75 & 0.89 \\
  \hline
  \multicolumn{2}{l}{$\lambda=1$} &  &  &  & &  &  &  & &  &  &  & \\
  0 & HD & 1421.70 & 1.00 & 3595.88 & 1.00 & 2296.03 & 1.00 & 1552.11 & 0.98 & 3258.39 & 1.00 & 2231.63 & 1.00 \\
  & 1S & 7121.11 & 1.00 & 6977.35 & 1.00 & 6976.16 & 1.00 & 7485.99 & 1.00 & 7182.76 & 1.00 & 7182.76 & 1.00 \\
  & LLA & 50.48 & 0.99 & 64.69 & 0.99 & 4.59 & 1.00 & 231.48 & 0.97 & 107.36 & 1.00 & 140.97 & 1.00 \\
  0.5 & HD & 4505.31 & 0.93 & 6764.71 & 0.88 & 4973.51 & 0.98 & 4571.62 & 0.97 & 6473.05 & 0.89 & 6150.70 & 0.96 \\
  & 1S & 11973.29 & 1.00 & 10301.59 & 1.00 & 10238.21 & 1.00 & 12411.82 & 1.00 & 9674.64 & 1.00 & 9781.43 & 1.00 \\
  & LLA & 1622.24 & 0.89 & 661.69 & 0.95 & 622.25 & 0.96 & 1682.66 & 0.89 & 1785.73 & 0.86 & 517.91 & 0.97 \\
  0.75$~$ & HD & 11166.35 & 0.75 & 16786.90 & 0.57 & 11642.59 & 0.84 & 12834.39 & 0.81 & 14964.11 & 0.66 & 10110.16 & 0.90 \\
  & 1S & 16953.51 & 1.00 & 9125.82 & 1.00 & 9225.76 & 1.00 & 17174.91 & 0.99 & 8828.81 & 1.00 & 8549.86 & 1.00 \\
  & LLA & 6379.56 & 0.50 & 4295.69 & 0.63 & 787.30 & 0.93 & 6904.11 & 0.52 & 3637.68 & 0.74 & 812.28 & 0.94 \\
  \hline
  \multicolumn{2}{l}{$\lambda=5$} &  &  &  & &  &  &  & &  &  &  & \\
  0 & HD & 12.35 & 1.00 & 12.35 & 1.00 & 12.35 & 1.00 & 13.00 & 1.00 & 13.00 & 1.00 & 13.00 & 1.00 \\
  & 1S & 1072.70 & 1.00 & 1072.70 & 1.00 & 1072.70 & 1.00 & 1114.13 & 1.00 & 1114.13 & 1.00 & 1114.13 & 1.00 \\
  & LLA & 0.01 & 1.00 & 0.05 & 1.00 & 0.01 & 1.00 & 0.01 & 1.00 & 0.07 & 1.00 & 0.01 & 1.00 \\
  0.5 & HD & 28.71 & 1.00 & 28.71 & 1.00 & 28.71 & 1.00 & 0.43 & 1.00 & 0.42 & 1.00 & 0.43 & 1.00 \\
  & 1S & 6793.73 & 1.00 & 6793.73 & 1.00 & 6793.73 & 1.00 & 6831.01 & 1.00 & 6831.01 & 1.00 & 6831.01 & 1.00 \\
  & LLA & 0.38 & 1.00 & 0.54 & 1.00 & 0.49 & 1.00 & 0.43 & 1.00 & 0.58 & 1.00 & 0.57 & 1.00 \\
  0.75$~$ & HD & 4998.08 & 0.88 & 4963.08 & 0.88 & 4292.65 & 0.97 & 5753.61 & 0.92 & 5772.76 & 0.95 & 5192.19 & 0.98 \\
  & 1S & 11191.83 & 1.00 & 11188.02 & 1.00 & 12029.12 & 1.00 & 11917.77 & 1.00 & 11971.47 & 1.00 & 12485.14 & 1.00 \\
  & LLA & 1217.39 & 0.90 & 1252.65 & 0.89 & 1060.08 & 0.99 & 861.72 & 0.95 & 937.76 & 0.94 & 1018.59 & 0.98 \\
  \hline
  \multicolumn{2}{l}{$\lambda=10$} &  &  &  & &  &  &  & &  &  &  & \\
  0 & HD & 0.00 & 1.00 & 0.00 & 1.00 & 0.00 & 1.00 & 0.00 & 1.00 & 0.00 & 1.00 & 0.00 & 1.00 \\
  & 1S & 0.00 & 1.00 & 0.00 & 1.00 & 0.00 & 1.00 & 0.00 & 1.00 & 0.00 & 1.00 & 0.00 & 1.00 \\
  & LLA & 0.00 & 1.00 & 0.00 & 1.00 & 0.00 & 1.00 & 0.00 & 1.00 & 0.00 & 1.00 & 0.00 & 1.00 \\
  0.5 & HD & 6.69 & 1.00 & 6.69 & 1.00 & 6.69 & 1.00 & 5.80 & 1.00 & 5.80 & 1.00 & 5.80 & 1.00 \\
  & 1S & 2883.52 & 1.00 & 2883.52 & 1.00 & 2883.52 & 1.00 & 2906.35 & 1.00 & 2906.35 & 1.00 & 2906.35 & 1.00 \\
  & LLA & 0.03 & 1.00 & 0.20 & 1.00 & 0.03 & 1.00 & 0.02 & 1.00 & 0.20 & 1.00 & 0.02 & 1.00 \\
  0.75$~$ & HD & 122.19 & 1.00 & 122.19 & 1.00 & 122.19 & 1.00 & 107.93 & 1.00 & 107.93 & 1.00 & 107.93 & 1.00 \\
  & 1S & 8835.88 & 1.00 & 8835.88 & 1.00 & 8835.87 & 1.00 & 8874.85 & 1.00 & 8874.85 & 1.00 & 8874.84 & 1.00 \\
  & LLA & 0.08 & 1.00 & 0.54 & 1.00 & 0.32 & 1.00 & 0.10 & 1.00 & 0.53 & 1.00 & 0.35 & 1.00 \\
  \hline
  \multicolumn{2}{l}{$\lambda=20$} &  &  &  & &  &  &  & &  &  &  & \\
  0 & HD & 0.00 & 1.00 & 0.00 & 1.00 & 0.00 & 1.00 & 0.00 & 1.00 & 0.00 & 1.00 & 0.00 & 1.00 \\
  & 1S & 0.00 & 1.00 & 0.00 & 1.00 & 0.00 & 1.00 & 0.00 & 1.00 & 0.00 & 1.00 & 0.00 & 1.00 \\
  & LLA & 0.00 & 1.00 & 0.00 & 1.00 & 0.00 & 1.00 & 0.00 & 1.00 & 0.00 & 1.00 & 0.00 & 1.00 \\
  0.5 & HD & 0.00 & 1.00 & 0.00 & 1.00 & 0.00 & 1.00 & 0.00 & 1.00 & 0.00 & 1.00 & 0.00 & 1.00 \\
  & 1S & 0.00 & 1.00 & 0.00 & 1.00 & 0.00 & 1.00 & 0.00 & 1.00 & 0.00 & 1.00 & 0.00 & 1.00 \\
  & LLA &  0.00 & 1.00 & 0.00 & 1.00 & 0.00 & 1.00 & 0.00 & 1.00 & 0.00 & 1.00 & 0.00 & 1.00 \\
  0.75$~$ & HD & 21.76 & 1.00 & 21.76 & 1.00 & 21.76 & 1.00 & 17.70 & 1.00 & 17.70 & 1.00 & 17.70 & 1.00 \\
  & 1S & 3997.88 & 1.00 & 3997.88 & 1.00 & 3997.88 & 1.00 & 4014.29 & 1.00 & 4014.30 & 1.00 & 4014.29 & 1.00 \\
  & LLA & 0.05 & 1.00 & 0.43 & 1.00 & 0.06 & 1.00 & 0.07 & 1.00 & 0.38 & 1.00 & 0.08 & 1.00 \\
   \hline
\end{tabular}}
\label{ngp.p81.sig13}
\end{center}
\end{table}

\subsection{Example 2: Binary Logistic Regression}

As in Example 1, we considered the LAS, ALAS, EN, AEN, and SCAD
penalties.  There are at least two \R packages that allow penalization
using the LAS and EN penalties: \textit{glmpath} \citep{glmpath}, which
handles binomial and poisson regression using a ``predictor-corrector''
method, and \textit{glmnet} \citep{glmnet}, which handles binomial and
multinomial regression using cyclical coordinate descent.  Both
methods can be tuned to find the same solutions, so for ease of
presentation we only consider the results of \textit{glmnet} for
comparison in the tables and analysis below.  The
\textit{SIS} package \citep{sis} permits computations with
the ALAS, AEN, and SCAD penalties using modifications of the
\cite{glmpath} code.  For SCAD, we compared the results of MIST to the
results from the one-step (1S) algorithm \citep[GLM version,
][]{zou:li} using the code provided from the authors and the LLA
algorithm as implemented in
\cite{sis}.

As before, we only considered comparing solutions that use the same
combination of tuning parameters; for the present example, the
set considered here is $\Lambda=\{0.001,0.01,0.05,0.1,0.2,1.00\}$,
reflecting a need to accommodate the different scaling of the
problem. The data generation scheme for this example was loosely based
on the simulation study found in \cite{glmnet}. Binary response data were
generated according to a logistic (rather than linear) regression
model using $p_i=[1+\exp(-\textbf{x}_i'\bfbeta^\ast)]^{-1}$,
$i=1,\ldots,N=1000$, where $\bfbeta^\ast$ is a $p-$vector with
elements $\beta_j=3\times(-1)^j\exp(-2(j-1)/200), ~j=1,\ldots,q,$
$q\in\{25,75\},$ and the remaining $100-q$ components set to zero.
Here, $\textbf{x}_i$ follows a $p$-dimensional multivariate normal
distribution with zero mean and covariance
$\boldsymbol{\Sigma}=3^{-2}\textbf{P}$ where correlation matrix
$\textbf{P}$ is such that each pair of predictors has the same
population correlation $\rho.$ We considered three such correlations,
$\rho \in \{0.0,0.5, 0.75\}.$

For the $B=100$ simulations, the maximum (across different tuning
parameters) average normed difference between the existing and
proposed methods, multiplied by $10^5,$ are reported for each of the
strictly convex cases in the right-most panel of
Table{~\ref{ngp:match:mn}}.  As before, these maximums are generally
remarkably small, indicating that MIST can recover the same (unique)
solutions as the existing algorithms.  The results for SCAD are
reported in Table{~\ref{ngp.q2575}}, which displays the same
information as in the corresponding tables from Example 1; the HD
comparisons are omitted here as the methodology and code were only
developed for the case of penalized least-squares.  In the GLM
setting, the 1S estimator is computed by applying the LARS
\citep{LARS} algorithm to a quadratic approximation of the negative
loglikelihood function evaluated at the MLE.  Thus, 1S takes `one
step' towards minimizing the objective function; in contrast, both
MIST and LLA iterate until a stationary point, usually a local
minimizer, is found. As in the linear model case, LLA uses
\textit{glmpath} to minimize the surrogate at each step, whereas the
MIST algorithm uses a single application of the soft thresholding
operator to minimize the surrogate at each step.

In this example, the starting value carried even greater importance in
comparison with the linear model setting. In particular, in the case
of MIST, the combination of a $\textbf{0}_p$ starting value and small
penalty parameter led to solutions with objective function evaluations
that were substantially larger in comparison with those obtained using
either $\widehat{\bfbeta}_{ml}$ and $\widehat{\bfbeta}_{1S,\lambda}.$
Such behavior may be directly attributed to the fact that the ML and
1S starting values either minimize or nearly minimize the negative
loglikelihood portion of the objective function, the dominant term in
the objective function when $\lambda$ is ``small.''  In contrast, a
$\textbf{0}_p$ starting value led to the best minimization performance
for ``large'' $\lambda$; upon reflection, this is also not very
surprising, since large penalties induce greater sparsity and
$\textbf{0}_p$ is the sparsest possible solution.

There were a few settings in which the
1S estimator resulted in a lower objective function evaluation in
comparison with applying MIST started at
$\widehat{\bfbeta}_{ml}$. This reflects the fact that several local
minima can exist for non-convex penalties like SCAD.  In addition, and
as was observed before, using the 1S solution as a starting value
always led to MIST finding a solution with a lower evaluation of the
objective function in comparison with that provided by the 1S
solution.  Regarding the use of LLA, which also requires a starting
value specification, we again examined the cases for which LLA resulted
in lower objective function evaluations.  For these cases, all MIST solutions
were LLA solutions, and all LLA solutions were MIST solutions with the exception of one.
Hence, both methods find valid, if often different, solutions, a behavior
that we again attribute to the differences in paths taken towards a
solution.

\begin{table}[!t]
\begin{center}
\caption{{\small Algorithm performance in Example 2 (GLM) for SCAD penalty. The column `avg' is the average normed differences ($\times 10^3$) between the MIST solution and the existing method's solution ; `prop' is the proportion of MIST solutions whose objective function evaluation was less than or equal to that of the existing method's solution.}}
{\scriptsize
\begin{tabular}{@{}llrr|rr|rr||rr|rr|rr@{}}
  \hline  \hline
        & & \multicolumn{6}{c||}{$q=25$} & \multicolumn{6}{c}{$q=75$} \\ \cline{3-14}
	$\bfbeta^{(0)}$  & & \multicolumn{2}{c|}{$\textbf{0}_p$} & \multicolumn{2}{c|}{$\widehat{\bfbeta}_{ml}$}
	& \multicolumn{2}{c||}{$\widehat{\bfbeta}_{1S,\lambda}$} & \multicolumn{2}{c|}{$\textbf{0}_p$} & \multicolumn{2}{c|}{$\widehat{\bfbeta}_{ml}$}
	& \multicolumn{2}{c}{$\widehat{\bfbeta}_{1S,\lambda}$}\\
	\cline{3-14}
\multicolumn{2}{l}{$\rho~~$ method} & avg & prop & avg & prop & avg & prop & avg & prop & avg & prop & avg & prop \\
  \hline
  \multicolumn{2}{l}{$\lambda=.001$} &  &  &  & &  &  &  & &  &  &  & \\
0 & 1S & 26.50 & 0.27 & 0.39 & 1.00 & 0.39 & 1.00 & 31.70 & 0.42 & 0.22 & 1.00 & 0.18 & 1.00 \\
& LLA & 18.55 & 0.68 & 0.15 & 1.00 & 0.13 & 1.00 & 17.31 & 0.76 & 0.22 & 1.00 & 0.11 & 1.00 \\
0.5 & 1S& 33.90 & 0.15 & 0.08 & 1.00 & 0.07 & 1.00 & 35.43 & 0.26 & 0.10 & 1.00 & 0.07 & 1.00 \\
& LLA & 27.65 & 0.64 & 0.01 & 1.00 & 0.00 & 1.00 & 18.45 & 0.82 & 0.10 & 1.00 & 0.00 & 1.00 \\
0.75 & 1S& 56.29 & 0.04 & 0.06 & 1.00 & 0.05 & 1.00 & 42.85 & 0.23 & 0.05 & 1.00 & 0.04 & 1.00 \\
& LLA & 46.48 & 0.71 & 0.05 & 1.00 & 0.00 & 1.00 & 26.05 & 0.82 & 0.04 & 1.00 & 0.00 & 1.00 \\
\hline
  \multicolumn{2}{l}{$\lambda=.01$} &  &  &  & &  &  &  & &  &  &  & \\
0 & 1S& 945.60 & 0.11 & 30.65 & 1.00 & 31.42 & 1.00 & 1318.20 & 0.02 & 8.61 & 1.00 & 8.61 & 1.00 \\
& LLA & 416.15 & 0.64 & 5.49 & 0.93 & 1.86 & 0.99 & 406.62 & 0.72 & 0.98 & 1.00 & 0.49 & 1.00 \\
0.5 & 1S& 1082.65 & 0.00 & 23.60 & 1.00 & 22.97 & 1.00 & 1088.23 & 0.01 & 5.62 & 1.00 & 5.75 & 1.00 \\
& LLA & 427.10 & 0.72 & 1.33 & 0.99 & 0.03 & 1.00 & 398.05 & 0.74 & 0.56 & 0.99 & 0.16 & 1.00 \\
0.75 & 1S& 1462.74 & 0.00 & 16.81 & 0.98 & 17.37 & 1.00 & 1629.73 & 0.00 & 5.53 & 0.99 & 4.97 & 1.00 \\
& LLA & 548.07 & 0.79 & 1.71 & 0.97 & 0.82 & 1.00 & 578.09 & 0.79 & 1.73 & 0.99 & 0.06 & 1.00 \\
\hline
  \multicolumn{2}{l}{$\lambda=.05$} &  &  &  & &  &  &  & &  &  &  & \\
0 & 1S& 1845.64 & 0.99 & 501.45 & 1.00 & 530.14 & 1.00 & 9575.27 & 0.82 & 252.36 & 1.00 & 263.41 & 1.00 \\
& LLA & 75.94 & 0.99 & 93.46 & 0.73 & 76.33 & 0.98 & 97.80 & 0.91 & 27.73 & 0.96 & 13.86 & 0.99 \\
0.5 & 1S& 4351.14 & 0.33 & 433.10 & 1.00 & 473.27 & 1.00 & 8323.46 & 0.98 & 171.08 & 1.00 & 181.11 & 1.00 \\
& LLA & 394.16 & 0.60 & 125.51 & 0.74 & 74.17 & 0.94 & 106.69 & 0.87 & 15.59 & 0.96 & 9.10 & 1.00 \\
0.75 & 1S& 5041.69 & 0.97 & 359.74 & 1.00 & 379.26 & 1.00 & 7907.54 & 1.00 & 156.65 & 0.99 & 164.34 & 1.00 \\
& LLA & 337.48 & 0.90 & 124.48 & 0.67 & 46.58 & 0.91 & 24.37 & 0.98 & 31.31 & 0.95 & 2.19 & 1.00 \\
\hline
  \multicolumn{2}{l}{$\lambda=.1$} &  &  &  & &  &  &  & &  &  &  & \\
0 & 1S& 4095.33 & 1.00 & 818.64 & 1.00 & 815.48 & 1.00 & 8626.86 & 1.00 & 834.01 & 1.00 & 832.92 & 1.00 \\
& LLA & 0.00 & 1.00 & 0.04 & 1.00 & 15.14 & 1.00 & 0.00 & 1.00 & 73.78 & 0.89 & 149.55 & 0.98 \\
0.5 & 1S& 4330.64 & 1.00 & 660.87 & 1.00 & 682.83 & 1.00 & 7626.58 & 1.00 & 628.29 & 1.00 & 718.12 & 1.00 \\
& LLA &  4.56 & 1.00 & 32.36 & 0.93 & 34.80 & 0.99 & 0.00 & 1.00 & 115.84 & 0.85 & 121.60 & 0.98 \\
0.75 & 1S& 4536.24 & 1.00 & 626.38 & 1.00 & 693.65 & 1.00 & 7457.80 & 1.00 & 550.76 & 1.00 & 618.94 & 1.00 \\
& LLA  & 0.00 & 1.00 & 81.21 & 0.87 & 87.10 & 0.99 & 0.00 & 1.00 & 88.95 & 0.86 & 62.41 & 0.98 \\
\hline
  \multicolumn{2}{l}{$\lambda=.2$} &  &  &  & &  &  &  & &  &  &  & \\
0 & 1S& 3712.07 & 1.00 & 2888.10 & 0.81 & 3712.07 & 1.00 & 4346.96 & 1.00 & 4346.96 & 1.00 & 4346.96 & 1.00 \\
& LLA & 0.00 & 1.00 & 0.04 & 1.00 & 0.01 & 1.00 & 0.00 & 1.00 & 0.01 & 1.00 & 0.01 & 1.00 \\
0.5 & 1S& 3768.77 & 1.00 & 3167.21 & 0.98 & 3602.53 & 1.00 & 3781.29 & 1.00 & 3781.29 & 1.00 & 3781.29 & 1.00 \\
& LLA & 0.00 & 1.00 & 42.80 & 0.99 & 70.75 & 1.00 & 0.00 & 1.00 & 0.01 & 1.00 & 0.01 & 1.00 \\
0.75 & 1S& 3825.82 & 1.00 & 2542.80 & 0.97 & 3076.24 & 1.00 & 4331.74 & 1.00 & 4331.74 & 1.00 & 4331.74 & 1.00 \\
& LLA & 0.00 & 1.00 & 404.72 & 0.83 & 387.72 & 0.86 & 0.00 & 1.00 & 0.01 & 1.00 & 0.01 & 1.00 \\
\hline
  \multicolumn{2}{l}{$\lambda=1$} &  &  &  & &  &  &  & &  &  &  & \\
0 & 1S& 54.18 & 1.00 & 54.18 & 1.00 & 54.18 & 1.00 & 61.54 & 1.00 & 61.54 & 1.00 & 61.54 & 1.00 \\
& LLA & 0.00 & 1.00 & 0.01 & 1.00 & 0.00 & 1.00 & 0.00 & 1.00 & 0.02 & 1.00 & 0.00 & 1.00 \\
0.5 & 1S& 40.38 & 1.00 & 40.38 & 1.00 & 40.38 & 1.00 & 49.01 & 1.00 & 49.01 & 1.00 & 49.01 & 1.00 \\
& LLA & 0.00 & 1.00 & 0.01 & 1.00 & 0.00 & 1.00 & 0.00 & 1.00 & 0.00 & 1.00 & 0.00 & 1.00 \\
0.75 & 1S& 32.85 & 1.00 & 32.85 & 1.00 & 32.85 & 1.00 & 38.36 & 1.00 & 38.36 & 1.00 & 38.36 & 1.00 \\
& LLA & 0.00 & 1.00 & 0.01 & 1.00 & 0.00 & 1.00 & 0.00 & 1.00 & 0.00 & 1.00 & 0.00 & 1.00 \\
   \hline
\end{tabular}}
\label{ngp.q2575}
\end{center}
\end{table}

\subsection{Effectiveness of SQUAREM$^2$}
\label{sec: cvg acc sim}

We explored the effectiveness of SQUAREM$^2$, defined in Section
\ref{sec: cvg acc}, when applied to several simulated datasets taken
from the previous two simulation studies.  Table{~\ref{spd}} indicates
the relative reduction in elapsed time (`RRT') and numbers of MM
updates, i.e., invocations of mapping $M(\cdot)$, required for the
original and SQUAREM$^2$-accelerated algorithms to converge for five
randomly chosen simulation datasets across the five penalty functions.
The SQUAREM$^2$ algorithm converged without difficulty in these cases
and required substantially fewer MM updates than the original
algorithm; the percent reduction in time was as high as 96\%. We
remark here that the regularity conditions imposed in \cite{rol:var}
and \cite{var:rol}, particularly smoothness conditions, are not
satisfied in this particular class of examples. Hence, while the
simulation results are certainly very promising, the question of
convergence (and its associated rate) of SQUAREM$^2$ in this class of
problems continues to remain an interesting open problem.

\begin{table}[!t]
\begin{center}
\caption{{\small Acceleration from SQUAREM$^2$ applied to simplified MIST algorithm for five randomly selected simulation datasets. The reduction in elapsed time is given by `RRT', while the number of MM updates are given for the original MIST implementation and SQUAREM$^2$ implementation in `\# orig' and `\# sqm$^2$', respectively.}}
{\scriptsize
\begin{tabular}{@{}r|l@{}r@{}r|l@{}r@{}r|l@{}r@{}r|l@{}r@{}r|l@{}r@{}r@{}}
  \hline \hline
 &\multicolumn{3}{c|}{LAS}&\multicolumn{3}{c|}{ALAS} &\multicolumn{3}{c|}{EN}&\multicolumn{3}{c|}{AEN}  &\multicolumn{3}{c}{SCAD} \\
Dataset & RRT & ~~\#orig  & ~\#sqm$^2$ & RRT & ~~\#orig & ~\#sqm$^2$  &  RRT  & ~~\#orig & ~\#sqm$^2$ &  RRT  & ~~\#orig & ~\#sqm$^2$ & RRT & ~~~\#orig & ~\#sqm$^2$ \\
  \hline
$LM$  &  &  & & & & & &  & &  &  & & &  &  \\
$p=35,\sigma=1$  &  &  & & & & & &  & &  &  & & &  &  \\
 62 & 0.67 & 260 &  62 & 0.81 & 169 &  44 & 0.63 &  46 &  26 & 0.82 &  42 &  23 & 0.91 & 485 &  68 \\
 71 & 0.76 & 221 &  59 & 0.75 & 163 &  41 & 0.67 &  49 &  29 & 0.62 &  44 &  29 & 0.83 & 302 &  65 \\
 86 & 0.67 & 271 &  68 & 0.70 & 149 &  44 & 0.67 &  51 &  29 & 0.75 &  43 &  26 & 0.93 & 987 & 104 \\
 95 & 0.86 & 317 &  74 & 0.88 & 187 &  41 & 0.92 &  49 &  29 & 0.73 &  46 &  26 & 0.90 & 500 &  71 \\
 88 & 0.88 & 330 &  68 & 0.87 & 162 &  41 & 0.78 &  51 &  29 & 0.77 &  45 &  26 & 0.90 & 528 &  77 \\
 $p=81,\sigma=3$&  &  & & & & & &  & &  &  & & &  &  \\
  62 & 0.90 & 2059 & 242 & 0.89 & 589 &  92 & 0.65 &  68 &  35 & 0.75 &  64 &  29 & 0.88 & 594 & 101 \\
  71 & 0.93 & 1426 & 164 & 0.93 & 838 &  83 & 0.76 &  77 &  32 & 0.70 &  71 &  32 & 0.94 & 2608 & 215 \\
  86 & 0.90 & 1351 & 212 & 0.92 & 956 &  98 & 0.59 &  77 &  38 & 0.79 &  69 &  32 & 0.92 & 1038 & 110 \\
  95 & 0.93 & 1500 & 167 & 0.86 & 367 &  71 & 0.67 &  72 &  35 & 0.74 &  68 &  29 & 0.90 & 663 &  92 \\
  88 & 0.92 & 1547 & 185 & 0.90 & 716 & 101 & 0.60 &  70 &  32 & 0.68 &  66 &  32 & 0.92 & 1798 & 203 \\
\hline
$GLM$  &  &  & & & & & &  & &  &  & & &  &  \\
$q=25$  &  &  & & & & & &  & &  &  & & &  &  \\
  62 & 0.93 & 4928 & 431 & 0.96 & 6227 & 272 & 0.89 & 3201 & 359 & 0.93 & 3316 & 236 & 0.95 & 22044 & 1442 \\
  71 & 0.92 & 4195 & 416 & 0.95 & 5045 & 239 & 0.90 & 2796 & 281 & 0.94 & 2843 & 170 & 0.95 & 16225 & 1052 \\
  86 & 0.92 & 4488 & 470 & 0.95 & 5449 & 242 & 0.92 & 2971 & 257 & 0.93 & 3044 & 206 & 0.95 & 20133 & 1193 \\
  95 & 0.93 & 4553 & 374 & 0.94 & 5419 & 341 & 0.92 & 3059 & 269 & 0.95 & 3096 & 152 & 0.95 & 15250 & 1064 \\
  88 & 0.92 & 5212 & 527 & 0.95 & 6850 & 371 & 0.91 & 3237 & 314 & 0.94 & 3393 & 203 & 0.96 & 26477 & 1367 \\
$q=75$  &  &  & & & & & &  & &  &  & & &  &  \\
  62 & 0.88 & 4334 & 674 & 0.91 & 3573 & 377 & 0.85 & 3055 & 575 & 0.90 & 2435 & 293 & 0.95 & 88994 & 5687 \\
  71 & 0.91 & 3805 & 446 & 0.92 & 3046 & 281 & 0.85 & 2761 & 536 & 0.89 & 2194 & 281 & 0.94 & 82615 & 5588 \\
  86 & 0.87 & 3615 & 602 & 0.91 & 2900 & 329 & 0.87 & 2653 & 434 & 0.92 & 2110 & 185 & 0.93 & 42652 & 3686 \\
  95 & 0.89 & 3870 & 554 & 0.90 & 3121 & 380 & 0.90 & 2820 & 338 & 0.89 & 2264 & 314 & 0.94 & 40002 & 3095 \\
  88 & 0.88 & 4177 & 641 & 0.94 & 3395 & 251 & 0.87 & 2972 & 482 & 0.91 & 2415 & 242 & 0.94 & 77484 & 5885 \\
   \hline \hline
\end{tabular}}
\label{spd}
\end{center}
\end{table}

\section{Example: Identifying genes
associated with the survival of lymphoma patients}\label{data}

Diffuse large-B-cell lymphoma (DLBCL) is an aggressive type of
non-Hodgkins lymphoma and is one of the most common forms of lymphoma
occurring in adults. \cite{rosetal} utilized Lymphochip DNA
microarrays, specialized to include genes known to be preferentially
expressed within the germinal centers of lymphoid organs, to collect
and analyze gene expression data from 240 biopsy samples of DLBCL
tumors. For each subject, 7399 gene expression measurements were
obtained. The expression profiles along with corresponding patient
information can be downloaded from their supplemental website
http://llmpp.nih.gov/DLBCL/.  Since the original profiles had some
missing expression measurements, we used the dataset subsequently
analyzed by \cite{li:gui} which estimated the missing values using a
nearest neighbor approach.  During the time of followup, 138 patient
deaths were observed with median death time of 2.8 years.

\cite{rosetal} used hierarchical clustering to group the genes into
four gene-expression signatures: Proliferation (PS), which includes
cell-cycle control and checkpoint genes, and DNA synthesis and
replication genes; Major Histocompatibility Complex ClassII (MHC),
which includes genes involved in antigen presentation; Lymph Node
(LNS), which includes genes encoding for known markers of monocytes,
macrophages, and natural killer cells; and Germinal Center B (GCB),
which includes genes that are characteristic of germinal center B
cells; see \cite{aliz:et:al} for more information on gene signatures.
Based on the gene clusters, they built a Cox proportional hazards
model \citep{Cox1972,
Cox1975}
to predict survival outcomes in the DLBCL patients.
Subsequently, this dataset has been analyzed numerous times, typically
to evaluate methods related to subgroup identification and/or survival
prediction \citep[e.g.,][]{li:gui, gui:li, guili:pen,
  li:lua, annetal, eng:li, cus}.

Here, we instead focus on the performance of two different penalties,
namely SCAD and MCP, with regard to the identification of genes
associated with DLBCL survival.  The simulation results of \cite{mcp}
suggest that MCP has superior selective accuracy over the SCAD
penalty, at least for the case of a linear model.  There, selection
accuracy was measured as the proportion of simulation replications
with correct classification of both the zero and non-zero
coefficients, with MCP outperforming SCAD in all simulation settings.
%
%%%%% In fact: it is the combination of MCP and the "PLUS" algorithm
%%%%% that is being summarized in these simulation results [and the PLUS 
%%%%% algorithm is also used to implement the SCAD penalty]. 
%
%%%%% In Zhang 2007 the following statement is made:
%
%%%%% MCP ensures the continuity and unbiasedness of sparse local
%%%%% minimizers of [the penalized least squares problem] to the
%%%%% greatest extent for general design matrices $X$
%
To illustrate the utility and flexibility of the MIST algorithm, we
reanalyzed the DLBCL data, fitting a penalized Cox regression model
respectively using SCAD and MCP penalty functions, and running these
procedures in combination with the Iterative Sure Independence
Screening procedure \citep[ISIS,][]{fan:sam:wu09} in order to ensure
that the dimension of the parameter space was maintained at a
manageable level.  For SCAD, we considered both the 1S and LLA
estimators.  The existing optimization functions provided in the
\textit{SIS} package for the ISIS procedure were used for the 1S
estimator, whereas relevant modifications to the ISIS code were made
in order to accommodate the fully iterative LLA and MCP estimators.
Optimization at each step of the ISIS algorithm in the case of the MCP
penalty utilized the MIST algorithm, as we are aware of no other
algorithm capable of fitting the Cox regression model subject to MCP
penalization.  The default settings in the \textit{SIS} package were
used to determine the maximum number of predictors ($[\frac{n}{4\log
  n}]=10$) and to define the additional ISIS parameters (e.g., use of
the unpenalized MLE as a starting value, ranking method, tuning
parameter selection) for all three analyses (1S-SCAD, LLA-SCAD,
MIST-MCP).  The parameter $a$ was set to 3.7 for all analyses; hence,
only the selection of $\lambda$ required any tuning.

Table{~\ref{DLBCL10}} displays the 11 genes identified by at least one
of the three analyses. The x's in a given column indicate the genes
with non-zero coefficients resulting from the corresponding
penalization.  The final column provides references for genes
previously linked to DLBCL in the literature.  Genes belonging to the
original \cite{rosetal} gene expression signatures are indicated with
parenthetical initials.  Note that the references provided are not
meant to be an exhaustive list, but instead to demonstrate the
relevance of certain genes and/or their altered expression levels in
DLBCL survival.

Interestingly, the LLA-SCAD and MIST-MCP penalizations selected the
same subset of genes, with a nearly a complete overlap with those
selected from the 1S-SCAD penalization. The number of genes selected
in each case is 10, the maximum specified by ISIS; 9 of these were
shared across the three penalizations.  According to NCBI Entrez Gene
search (http://www.ncbi.nlm.nih.gov/), many of these genes are
biologically relevant.  For example, CDK7 codes for a protein that
regulates cell cycle progression and is represented in the
Proliferation Signature, although reported under a different
Lymphochip ID as this gene was spotted multiple times on the array.
Also members of the Proliferation Signature are SEPT1, coding for a
protein involved in cytokinesis, and BUB3, coding for a mitotic
checkpoint protein.  DNTTIP2 regulates transcriptional activity of
DNTT, a gene for a protein expressed in a restricted population of
normal and malignant pre-B and pre-T lymphocytes during early
differentiation.  HLA-DRA, a member of the MHC Signature, plays a
central role in the immune system and is expressed in antigen
presenting cells, such as B lymphocytes, dendritic cells, macrophages.
From the GCB Signature, the ESTs weakly similar to thyroxine-binding
globulin precursor is highly cited.  Additionally, RFTN1 plays a
pivotal role in regulating B-cell antigen receptor-mediated signaling
\citep{raftlin}.

A description of AI568329 was not provided in the original dataset,
thus its function is unknown.  Similarly, although cited at least
twice, a description for AA830781 was also not provided in the
original dataset.  However, both of these may be related to lymphoma
or risk of death from lymphoma, as it is possible that these genes
(and potentially others) were selected because of coexpression or
correlation with other relevant genes.

Interestingly the two genes not commonly identified across the three
penalizations were both cited in \cite{mart:et:al}.  They found
altered gene expression of TSC22D3 and ITGAL (both involved in a
variety of immune phenomena) in one case who initially presented with
follicle center lymphoma and subsequently transformed to DLBCL.

\begin{table}[!th]
\begin{center}
\caption{{\small Genes associated with DLBCL survival with SCAD (one-step=1S and LLA) and MCP penalizations, sorted by the gene order in
the original data set.  ID refers to the unique Lymphochip identification number.  The x's in a given column indicate the genes identified by the corresponding penalization.}}
{\scriptsize
\begin{tabular}{@{}lll@{}c@{}c@{}l@{}}
  \hline
ID	&	Name (Symbol)	& \multicolumn{2}{c}{SCAD} & MCP & ~References\\
 &  & 1S$~~$ & LLA$~~$ &  &  \\
\hline \hline
27774	& cyclin-dependent kinase 7 (CDK7)					&	x	&	x	&	x & \cite{rosetal} (PS), \cite{ma:huang} \\
& & & & & \cite{bin:sch08,bin:sch} \\ \hline
31242	& acidic 82 kDa protein mRNA (DNTTIP2)			&	x	&	x	&	x & \cite{bin:sch08,bin:sch} \\ \hline
31981	& septin 1 (SEPT1)													&	x	&	x	&	x & \cite{rosetal} (PS), \cite{li:lua} \\
& & & & & \cite{sin:neu:van}, \cite{shaetal} \\
& & & & & \cite{zha2},\cite{annetal} \\
& & & & & \cite{bin:sch08,bin:sch}\\ \hline
29652	& BUB3 budding uninhibited by benzimidazoles 3	(BUB3)	
																									&	x	&	 x	&	x & \cite{rosetal} (PS)\\ \hline
27731	& major histocompatibility complex, 				&	x	&	x	&	x & \cite{rosetal} (MHC),\cite{li:lua} \\
& class II, DR alpha (HLA-DRA) & & & & \cite{guili:pen,gui:li},\cite{sohn:et:al} \\
& & & & & \cite{bin:sch}\\ \hline
24376	& ESTs, Weakly similar to A47224  					& 	x	&	x	&	x & \cite{rosetal} (GCB),\cite{ando:et:al} \\
			& thyroxine-binding globulin precursor & & & &  \cite{guili:pen,gui:li},\cite{li:lua}\\	
			& & & & & \cite{annetal}, \cite{sohn:et:al}\\
			& & & & & \cite{bin:sch08,bin:sch}\\
			\hline
22162	& delta sleep inducing peptide, immunoreactor (TSC22D3)	
																									&		&	 x	&	x & \cite{mart:et:al} \\ \hline
23862	& (AI568329) ESTs														&	x	&	x	&	x & \\ \hline

24271 & integrin, alpha L	(ITGAL)									& x 	& 	& 		& \cite{mart:et:al} \\ \hline
33358 & (AA830781)																&	x	&	x	& x & \cite{li:lua} \\			
& & & & & \cite{bin:sch} \\ \hline
32679	& KIAA0084 protein (RFTN1)									&	x	&	x	&	x & \cite{gui:li}, \cite{shaetal}\\
& & & & & \cite{zha2},\cite{annetal}\\		
& & & & & \cite{bin:sch08,bin:sch} \\
\hline
\end{tabular}}
\label{DLBCL10}
\end{center}
\end{table}

The results of this analysis demonstrate equivalence in selection
performance between MCP and LLA-SCAD for the case of Cox proportional
hazards model.  Increasing the maximum number of predictors to $21$
again resulted in equivalent selection performance between MCP and
LLA-SCAD, with 21 predictors ultimately selected (results not
shown). The 1S estimator also resulted in the selection of 21
predictors, but with increased dissimilarity between MCP/LLA-SCAD and
1S: only 13 of the 21 genes were selected by all three methods.  It
should be noted that \cite{mcp} did not use any form iterative
variable selection (e.g., ISIS) in his comparisons between SCAD and
MCP for the case of the linear model; in addition, \cite{mcp} fixed
values for both penalty parameters in his simulations and also did not
use $a=3.7$.  Thus, use of the ISIS procedure, the particular method
used for selecting $\lambda$, and the use of $a=3.7$ (as suggested in
\cite{fan:sam:wu09}) in both the MCP and SCAD penalties may all play a
role in the results summarized above.

\section{Discussion}

This paper proposed a versatile and general algorithm capable of
dealing with a wide variety of nonsmoothly penalized objective
functions, including but not limited to all presently popular
combinations of data fidelity and penalty functions. We established a
suitable convergence theory, as well as new results on the convergence
of general MM algorithms. We also demonstrated the remarkable
effectiveness of the SQUAREM$^2$ acceleration procedure in these
problems as tool for accelerating the slow but steady convergence of
the proposed class of MM algorithms.  Beyond specification of the
penalty parameter(s) $\bflam,$ virtually no effort was expended in
tuning or otherwise specializing the MIST algorithm for solving a
given problem. Thus, at the expense of greater analytical work, the
convergence rate of the MIST algorithm can likely be improved. Through
the use relaxation techniques and other methods for controlling the
step-size behavior (e.g., line-searches) of MIST, we further expect
that the local nature of the convergence theory presented here can be
made global in nature.

The simulation results of this paper highlight the fact that nonconvex
penalties tend to endow the corresponding objective function with
multiple local minima.  The resulting sensitivity of computational
algorithms to the choice of starting value, while known, has arguably
been deemphasized in the current literature.  In this regard, the
one-step method of \cite{zou:li} provides a meritorious choice of
starting value for fully iterative SCAD-based algorithms. In addition
to being unique under mild regularity conditions, it is easily
generalized to other nonconvex penalties, such as MCP.  Unfortunately,
the utility of this approach for identifying starting values is also
limited to settings where $N>p$, for the justification of the 1S
estimator relies heavily on the use of the unpenalized MLE as its
starting value.

The simulated examples in this paper only consider settings with
$N>p$, mainly to ensure that $m(\bfbeta)$ is strictly convex.
Specifying $\epsilon>0$ in the ridge-like penalty term ensures that
$m(\bfbeta)$ is strictly convex provided only that $g(\bfbeta) +
h(\bfbeta,\bfalpha)$ is convex, as might be encountered in cases where
$N<p$.  Thus, for example, one might consider combining the ridge term
with any penalty satisfying condition (P1) (e.g., SCAD), providing
alternatives to the elastic net penalty; our results on the
convergence of the proposed algorithms to some stationary point of the
objective function would continue to apply in this setting.  It would
be interesting to investigate the statistical properties of estimators
derived under such combinations in settings where $p > N$ but $p_0 \ll
N$, with $p_0$ denoting the number of ``important'' predictors.

\bibliographystyle{ims}
\bibliography{bibfile}

\appendix

\section{Appendix}

This appendix is divided into several sections. Section
\ref{app:MMlocal} reviews and extends the convergence theory for the
EM algorithm established in \cite{Wu}; the extension utilizes results
of \cite{Meyer} to establish stronger convergence results for general
MM algorithms. Section \ref{app:genpenMM} contains the proof of
Theorem \ref{thm:gen_penal_MM} and makes direct use of these new
convergence results. Finally, Sections \ref{app:genMIST} and \ref{app:surglm}
respectively contain the proofs of Theorems \ref{thm: gen_MIST} and
\ref{prop:sur:glm}, establishing the convergence of iterated soft
thresholding when used to minimize \eqref{eq: MIST object 1} and
convergence of the proposed class of MIST algorithms in the case of
the generalized linear model.

\input{new_appendix.tex}

\input{main_MM_thm_proof.tex}

\input{thm_gen_MIST_proof.tex}

\input{glm_thm_proof.tex}

\end{document}

%% file: new_appendix.tex
%==================================================================
%==================================================================

\subsection{Local convergence of MM algorithms in nonsmooth problems}
\label{app:MMlocal}

Using convergence theory for algorithms derived from point-to-set maps
developed by \citet{Zangwill}, \citet{Wu} established the convergence
of the EM algorithm assuming twice differentiability of the
loglikelihood function. In what follows, we first restate the key
convergence result of \citet{Zangwill}; this result, given in Theorem
\ref{thm: Zangwill} and adapted from \citet{Wu}, is stated in a form
convenient for use with the MM algorithm and provides for a very
general (and comparatively weak) form of convergence.  We then draw on
stronger convergence results due to \citet{Meyer} in order to
establish a more useful convergence theory for MM algorithms designed
to minimize nondifferentiable objective functions; this result is
stated in Theorem \ref{thm: Meyer}. Finally, we provide a set of
sufficient regularity conditions that ensure the validity of the
conditions of both theorems in a wide class of statistical
estimation problems.

Let $\obfn(\bfbeta)$ be the real-valued function to be minimized,
where $\bfbeta \in {\cal B}$ and ${\cal B}$ is some convex subset of
$\Real^p$. Let $M: {\cal B} \rightarrow {\cal B}$ be the minimization
map \eqref{the map}, where $\sobfn(\cdot,\cdot)$ is any function that
majorizes $\obfn(\bfbeta)$ for $\bfbeta \in {\cal B}$. In general,
$M(\cdot)$ is a point-to-set map, and therefore a set.  We say that
$\bar{\bfbeta}$ is a generalized fixed point of $M(\cdot)$ if
$\bar{\bfbeta} \in M(\bar{\bfbeta})$; we say that $\bar{\bfbeta}$ is a
fixed point of $M(\cdot)$ if $M(\bar{\bfbeta}) = \{ \bar{\bfbeta} \}$
(i.e., a singleton).

The main result of \citet[Theorem A]{Zangwill}, also 
utilized in \citet{Wu}, is stated below. 

\begin{thm}
\label{thm: Zangwill}
Suppose $\obfn(\bfbeta)$ is a continuous, real-valued function of
$\bfbeta \in {\cal B}$ that is uniformly bounded below. Let
${\cal S} \subset {\cal B}$ denote the (nonempty) set of stationary
points of $\obfn(\bfbeta)$ for $\bfbeta \in {\cal B}$
%; let
%${\cal M} \subset {\cal S}$ denote the (nonempty) set of
%minimizers of $\obfn(\bfbeta)$ for $\bfbeta \in {\cal B}$.
and assume the sequence $\{\bfbeta^{(k)}, k \geq 0\}$ is generated as follows:
\begin{itemize}
\item $\bfbeta^{(0)} \in {\cal B},$ where
$\bfbeta^{(0)}$ and $\obfn(\bfbeta^{(0)})$ are bounded;

\item $\bfbeta^{(k+1)} \in M(\bfbeta^{(k)}),$ where
$M(\cdot)$ is the point-to-set map \eqref{the map}.
\end{itemize}

Suppose that
\begin{itemize}
\item[Z1.] Each $\bfbeta^{(k)} \in {\cal B}_0$, where
the compact set ${\cal B}_0 \subset {\cal B}$;

\item[Z2.] $M(\cdot)$ is closed and non-empty for $\bfbeta \in
{\cal S}^c \cap {\cal B}_0$.

\item[Z3.] We have:
\begin{itemize}
\item[(i)]  $\obfn(\bfbeta) \leq \obfn(\bfalpha)$ for each 
$\bfalpha \in {\cal S}$ and any $\bfbeta \in
  M(\bfalpha);$ 
\item[(ii)] $\obfn(\bfbeta) < \obfn(\bfalpha)$
for each $\bfalpha \not \in {\cal S}$ and any $\bfbeta \in
  M(\bfalpha).$
\end{itemize}

% \item For each  $\bfalpha \in \Gamma^c$ and any $\bfbeta \in A(\bfalpha),$
% $\obfn(\bfbeta) < \obfn(\bfalpha);$

% \item For each  $\bfalpha \in \Gamma$ and any $\bfbeta \in A(\bfalpha),
% $\obfn(\bfbeta) \leq \obfn(\bfalpha);$

% \item[Z3.] For each $\bfalpha \in {\cal B}_0$ and any $\bfbeta \in
%   A(\bfalpha),$ we have (i) $\obfn(\bfbeta) \leq \obfn(\bfalpha);$
%   and, (ii) $\bfbeta \in A(\bfalpha)$ and $\obfn(\bfbeta) =
%   \obfn(\bfalpha)$ implies $\bfbeta = \bfalpha$.

\end{itemize}

Then, the following conclusions hold:
\begin{itemize}
\item[M1.] The sequence $\{\bfbeta^{(k)}, k\geq 0\}$ has at least one
limit point in ${\cal S},$ and the set of all limit points,
say ${\cal S}_0$, satisfies ${\cal S}_0 \subseteq {\cal S};$

\item[M2.] Each limit point $\bar{\bfbeta} \in {\cal S}_0$ satisfies
$\lim_{k \rightarrow \infty} \obfn(\bfbeta^{(k)})
= \obfn(\bar{\bfbeta}).$

\item[M3.] Each limit point $\bar{\bfbeta} \in {\cal S}_0$ 
is a generalized fixed point of $M(\cdot)$.
\end{itemize}
\end{thm}

\begin{remark} Assumptions [Z1]-[Z3] are imposed in \citet{Wu}. The
assumption [Z1] implies that $\{\bfbeta^{(k)}, k \geq 0\}$ is a
bounded sequence, ensuring the existence of at least one limit point.
Further comments on [Z2] will be made below, as it is possible to
impose reasonable sufficient conditions that ensure this condition.
The assumption [Z3] enforces the descent property at each
update, as would be expected in any EM, GEM or MM algorithm.
An equivalent formulation of this condition follows
\citep[e.g.][p.\ 114]{Meyer}:
\begin{itemize}
\item[Z3$'$.] For each $\bfalpha \in {\cal B}_0$
and $\bfbeta \in M(\bfalpha):$

\begin{itemize}
\item[(i)] $\obfn(\bfbeta) < \obfn(\bfalpha)$ if
$\bfalpha \not \in M(\bfalpha)$ (i.e.,
a strict decrease occurs at points $\bfalpha$ 
that are not generalized fixed points);

\item[(ii)] $\obfn(\bfbeta) \leq \obfn(\bfalpha)$ if $\bfalpha \in M(\bfalpha)$ 
(i.e., if $\bfalpha$ is a generalized fixed point, 
it is possible to observe no change in the objective function).
\end{itemize}
\end{itemize}
\end{remark}

The above theorem essentially guarantees convergence of subsequences,
but not global convergence of the iteration sequence
itself. Subsequential convergence permits, for example, oscillatory
behavior in the limit sequence.  \citet{Meyer,Meyer77} offers several
refinements of Theorem \ref{thm: Zangwill}, strengthening the
statements of convergence. His results, adapted for the MM algorithm,
follow below; in particular, see Theorem 3.1, Corollary 3.2, and
Theorems 3.5 and 3.6 of \citet{Meyer}.
\begin{thm}
\label{thm: Meyer}
Let the conditions of Theorem \ref{thm: Zangwill} hold. 
Consider the following two additional conditions:
\begin{itemize}
\item[Z4.] For each $\bfalpha \in {\cal B}_0$ and any $\bfbeta \in
  M(\bfalpha),$ we have $\obfn(\bfbeta) < \obfn(\bfalpha)$ 
  whenever $M(\bfalpha) \neq \{\bfalpha \}$ (i.e.,
a strict decrease in the objective function occurs at
any point $\bfalpha$ that is not a fixed point);

\item[Z5.] there exists an isolated limit point $\bar{\bfbeta}^*$
such that $M(\bar{\bfbeta}^*) = \{\bar{\bfbeta}^*\}$ (i.e.,
a true fixed point).
\end{itemize}

% \item For each  $\bfalpha \in \Gamma^c$ and any $\bfbeta \in A(\bfalpha),$
% $\obfn(\bfbeta) < \obfn(\bfalpha);$

% \item For each  $\bfalpha \in \Gamma$ and any $\bfbeta \in A(\bfalpha),
% $\obfn(\bfbeta) \leq \obfn(\bfalpha);$

% \item[Z3.] For each $\bfalpha \in {\cal B}_0$ and any $\bfbeta \in
%   A(\bfalpha),$ we have (i) $\obfn(\bfbeta) \leq \obfn(\bfalpha);$
%   and, (ii) $\bfbeta \in A(\bfalpha)$ and $\obfn(\bfbeta) =
%   \obfn(\bfalpha)$ implies $\bfbeta = \bfalpha$.

Suppose [Z1]-[Z4] hold. Then, in addition to results [M1]-[M3]
of Theorem \ref{thm: Zangwill}, the following conclusions hold:
\begin{itemize}
\item[M4.] Each limit point $\bar{\bfbeta} \in {\cal S}_0$ satisfies
$M(\bar{\bfbeta}) = \{ \bar{\bfbeta} \}$,
and is therefore a fixed point of $M(\cdot)$;

\item[M5.] $\lim_{k \rightarrow \infty} \|\bfbeta^{(k)} -
  \bfbeta^{(k+1)} \| = 0,$ in which case one either has (i) the set of
  limit points ${\cal S}_0$ consists of a single point to which
  $\bfbeta^{(k)}$ converges; or, (ii) the set of limit points ${\cal
    S}_0$ forms a continuum, and $\bfbeta^{(k)}$ fails to converge;

\item[M6.] If the number of fixed points having any given value
of $\obfn(\cdot)$ is finite, then $\{ \bfbeta^{(k)}, k \geq 0\}$
converges to one of these fixed points;

\item[M7.] 
 If the sequence $\{\bfbeta^{(k)}, k\geq 0\}$ has an isolated
  fixed point $\bar{\bfbeta},$ then $\bfbeta^{(k)} \rightarrow
  \bar{\bfbeta}$. If $\bar{\bfbeta}$ is also an
  isolated local minimum of $\obfn(\cdot)$ on ${\cal B}_0,$
  then there exists an open neighborhood ${\cal B}_{\epsilon} \subseteq
  {\cal B}_0$ of $\bar{\bfbeta}$ such that $\bfbeta^{(k)} \rightarrow
  \bar{\bfbeta}$ if $\bfbeta^{(0)} \in {\cal B}_{\epsilon}.$
\end{itemize}

Suppose instead that [Z1-Z3] and [Z5] hold. Then, in addition to
results [M1]-[M3] of Theorem \ref{thm: Zangwill}, the following
conclusion can be drawn:
\begin{itemize}
\item[M8.] If the sequence $\{\bfbeta^{(k)}, k\geq 0\}$ has an
  isolated generalized fixed point $\bar{\bfbeta}$ that satisfies
  $M(\bar{\bfbeta}) = \{ \bar{\bfbeta} \},$ then $\bfbeta^{(k)}
  \rightarrow \bar{\bfbeta}$. If $\bar{\bfbeta}$ is also an isolated
  local minimum of $\obfn(\cdot)$ on ${\cal B}_0,$ then there exists an
  open neighborhood ${\cal B}_{\epsilon} \subseteq {\cal B}_0$ of
  $\bar{\bfbeta}$ such that $\bfbeta^{(k)} \rightarrow \bar{\bfbeta}$
  if $\bfbeta^{(0)} \in {\cal B}_{\epsilon}.$
\end{itemize}
\end{thm}

\begin{remark} Assumption [Z4] strengthens [Z3] by imposing the
condition that the iteration scheme is {\em strictly} monotonic; as
such, all generalized fixed points of $M(\cdot)$ are also fixed
points, a situation that permits stronger statements of convergence
results.  Assumption [Z5] imposes the somewhat weaker assumption that
there exists at least one isolated fixed point of the iteration
sequence; similarly to [M7], [M8] implies that the iteration converges
to this point.
\end{remark}

Conclusions [M1]-[M7] essentially mirror those in \cite[Theorems
1-3]{vaida}, who obtains strong convergence results for the EM and MM
algorithms under global differentiability assumptions on the objective
and majorization functions and the additional condition that
$\sobfn(\bfbeta,\bfalpha)$ has a unique global minimizer in $\bfbeta$
for each $\bfalpha \in {\cal S},$ where ${\cal S}$ is a finite set of
isolated stationary points. This uniqueness condition, encapsulated in
[Z4], provides a verifiable condition for convergence of
the MM algorithm that is often satisfied in statistical
applications.

%**************************

Sufficient conditions that ensure [Z1]-[Z4], but weaker than
conditions imposed in \citet{vaida}, are now provided. In particular,
suppose that the objective function, its surrogate and the mapping
$M(\cdot)$ satisfy the following regularity conditions:
\begin{itemize}
\item[R1.] $\obfn(\bfbeta)$ is locally Lipschitz continuous and
  coercive for $\bfbeta \in {\cal B}$; that is, $L(\obfn(\textbf z)) =
  \{\textbf b \in {\cal B}: \obfn(\textbf b) \leq \obfn(\textbf z) \}$
  is compact for each $\textbf z \in {\cal B}$. Consequently,
  $\obfn(\bfbeta)$ achieves a finite minimum somewhere interior to
  ${\cal B}$; assume the set of stationary points
  ${\cal S}$ is finite and isolated.

%  This implies that
%   $\obfn(\bfbeta)$ is lower semicontinuous and lower compact; it also
%   implies that $\obfn(\bfbeta)$ achieves a finite minimum somewhere
%   interior to $\Real^p$ and approaches $+\infty$ if, and only if,
%   $\|\bfbeta\| \rightarrow \infty$ \citep[][\S
%   10.3]{optim}.%(Lange, 2004, \S 10.3).

\item[R2.] $\obfn(\bfbeta) = \sobfn(\bfbeta,\bfbeta)$ for each
$\bfbeta \in {\cal B}$.

\item[R3.] $\sobfn(\bfbeta,\bfalpha) > \sobfn(\bfbeta,\bfbeta)$
for $\bfbeta \neq \bfalpha$,
$\bfbeta, \bfalpha \in {\cal B}.$

\item[R4.] $\sobfn(\bfbeta,\bfalpha)$ is 
continuous for $(\bfalpha, \bfbeta) \in {\cal B} \times {\cal B}$
and locally Lipschitz in $\bfbeta$ for $\bfbeta$ near $\bfalpha$.

% \textbf{why do we need this in $\bfalpha$ also? is just
% plain continuity enough?  See Vaida, p833, R5}

\item[R5.] $M(\bfbeta)$ exists and is 
a singleton set for each $\bfbeta \in
{\cal B}$.
%; moreover, $M(\bfbeta)$ is continuous at each $\bfbeta \in {\cal S}$.

% \item[R6.] The function $\psi(\bfbeta,\bfalpha) :=
%   \sobfn(\bfbeta,\bfalpha) - \obfn(\bfbeta)$ is locally Lipschitz
%   continuous in $\bfbeta$ for $\bfbeta$ near $\bfalpha$. Moreover, for
%   each $\bfalpha$, the Clarke subdifferential \citep{clarke} of
%   $\psi(\bfbeta,\bfalpha)$ in $\bfbeta$ reduces to the singleton
%   $\{\textbf 0\}$ at $\bfbeta = \bfalpha$.

%\item $\obfn(\bfbeta_{n+1}) \leq \obfn(\bfbeta_{n})$
\end{itemize}

The above conditions do not imply that the
objective function $\obfn(\bfbeta)$ is differentiable everywhere.
Condition R1 does imply that $\obfn(\bfbeta)$ is bounded for $\bfbeta$
interior to ${\cal B}$ and that $\nabla \obfn(\bfbeta)$ exists for
almost all $\bfbeta.$ Conditions R2 and R3 imply that
$\sobfn(\bfbeta,\bfalpha)$ strictly majorizes $\obfn(\bfbeta)$ and, in
addition, \benum
\label{surr}
\sobfn(\bfbeta,\bfalpha) = \obfn(\bfbeta) + \psi(\bfbeta,\bfalpha),
\eenum where $\psi(\bfbeta,\bfalpha) := \sobfn(\bfbeta,\bfalpha) -
\obfn(\bfbeta)$ satisfies $\psi(\bfbeta,\bfalpha) > 0$ for $\bfalpha
\neq \bfbeta$ and $\psi(\bfbeta,\bfbeta) = 0.$ Assumptions R4 \& R5
imply that the map $M(\bfbeta)$ is continuous, hence bounded on
compact sets \citep[][Prop. 3.2]{Polak87}.  Conditions R1, R4, and R5
further imply that \eqref{surr} is bounded below for
$(\bfalpha, \bfbeta) \in {\cal B} \times {\cal B}$ and that
$\psi(\bar{\bfbeta},\bfalpha)$ is uniquely minimized at $\bfalpha =
\bar{\bfbeta}$ for any fixed point $\bar{\bfbeta}.$

Suppose conditions R1-R5 hold. As commented earlier, conditions R4 and R5 
imply that $M(\bfbeta)$ is a continuous point-to-point map; hence,
$M(\cdot)$ is closed \citep[e.g.][pp.\ 203-204]{Luenberger08},
establishing [Z2]. Propositions \ref{prop A} and \ref{prop B}, given
below and proved under conditions R1-R5,
now establish [Z1], [Z3] and [Z4].
%In fact, Theorem 4.3 of \citet{Meyer} implies that [Z4] also holds.

\begin{prop}
\label{prop A}
Suppose $\bfbeta^{(n)}$ is bounded for a
given $n \geq 0$. Then, $\bfbeta^{(n+1)} = M(\bfbeta^{(n)})$ exists,
is bounded and is unique. In addition, for $n \geq 0,$
\benum
\label{surr ineq}
\sobfn(\bfbeta^{(n+1)},\bfbeta^{(n)}) \leq \sobfn(\bfbeta^{(n)},\bfbeta^{(n)}) < \infty
\eenum
and
\benum
\label{decreasing}
\obfn(\bfbeta^{(n+1)}) -
 \obfn(\bfbeta^{(n)}) \leq -\psi(\bfbeta^{(n+1)},\bfbeta^{(n)}) \leq 0,
\eenum
where the second inequality 
is strict unless $\bfbeta^{(n+1)} = M(\bfbeta^{(n)}) = 
\bfbeta^{(n)}$.\\
\end{prop}

\begin{prop}
\label{prop B}
Let $\bfbeta^{(0)}$ be bounded.
Define $\obfn^{(n)} = \obfn(\bfbeta^{(n)})$ for $n \geq 0$.
Then, $\{\obfn^{(n)}, n \geq 0\}$ is a bounded, monotone decreasing
sequence. Moreover, the sequence $\{ \bfbeta^{(n)}, n \geq 0\}$ is bounded
and contained in the compact set $L(\xi^{(0)})$.\\
\end{prop}

\noindent{\sf Proof of Proposition \ref{prop A}:} Let $\bfalpha$ be bounded but otherwise
arbitrary. The continuity of $M(\cdot)$, along with assumption R5,
implies that $M(\bfalpha)$ exists, is bounded, and is unique.
Using \eqref{the map} and Assumption R2, we have that
$\sobfn(M(\bfalpha),\bfalpha) \leq \sobfn(\bfalpha,\bfalpha) =
\obfn(\bfalpha) < \infty$. Hence, \eqref{surr ineq} holds upon setting
$\bfalpha = \bfbeta^{(n)}.$

To establish \eqref{decreasing}, note that \eqref{surr},
\eqref{surr ineq} and the definition of $\bfbeta^{(n+1)}$ imply
\[
\sobfn(\bfbeta^{(n+1)},\bfbeta^{(n)}) = \obfn(\bfbeta^{(n+1)}) +
\psi(\bfbeta^{(n+1)},\bfbeta^{(n)}) <
\infty.
\]
By \eqref{surr ineq} and the fact that
$\sobfn(\bfbeta^{(n)},\bfbeta^{(n)}) = \obfn(\bfbeta^{(n)}) +
\psi(\bfbeta^{(n)},\bfbeta^{(n)}) = \obfn(\bfbeta^{(n)}),$
we further observe
\[
\obfn(\bfbeta^{(n+1)}) + \psi(\bfbeta^{(n+1)},\bfbeta^{(n)})
\leq  \obfn(\bfbeta^{(n)}). 
\]
from which \eqref{decreasing} is immediate. 
Under R3 and R4, this inequality is necessarily
strict unless 
$\bfbeta^{(n+1)} = M(\bfbeta^{(n)}) = 
\bfbeta^{(n)}$, proving the result. $\Box$\\

\noindent{\sf Proof of Proposition \ref{prop B}:} 
Since $\bfbeta^{(0)}$ is bounded, Assumption R1 implies $\obfn^{(0)}$
is bounded, $\bfbeta^{(0)} \in L(\obfn^{(0)})$, and $L(\obfn^{(0)})$
is compact.  From Proposition \ref{prop A} and Assumption R5, we
further observe that $\bfbeta^{(1)} = M(\bfbeta^{(0)})$ is bounded and
satisfies $\bfbeta^{(1)} \in L(\obfn^{(0)})$. Using Assumption R1 once
more, $\obfn^{(1)} = \obfn(\bfbeta^{(1)})$ is bounded and, by
\eqref{decreasing}, satisfies $\obfn^{(1)} \leq \obfn^{(0)}$; thus,
$L(\obfn^{(1)}) \subset L(\obfn^{(0)}).$

We now use induction. Let $\bfbeta^{(n)}$ be bounded for some $n \geq
1$ and satisfy $\obfn^{(n)} \leq \obfn^{(0)}$; then, $\obfn^{(n)}$ is
necessarily bounded and $\bfbeta^{(n)} \in L(\obfn^{(n)}) \subset
L(\obfn^{(0)}).$ It again follows from Proposition \ref{prop A} and
Assumption R5 that $\bfbeta^{(n+1)} = M(\bfbeta^{(n)})$ is bounded and
satisfies $\bfbeta^{(n+1)} \in L(\obfn^{(n)})$. Hence, $\obfn^{(n+1)}$
is bounded and satisfies $\obfn^{(n+1)} \leq \obfn^{(n)} \leq
\obfn^{(0)}$.  Consequently, we have $L(\obfn^{(n+1)}) \subset
L(\obfn^{(n)}) \subset L(\obfn^{(0)})$ and $\bfbeta^{(n+1)} \in
L(\obfn^{(0)})$ and it now follows that $\obfn^{(n+1)} \leq
\obfn^{(n)}$, $L(\obfn^{(n+1)}) \subset L(\obfn^{(n)}) \subset
L(\obfn^{(0)})$, and $\bfbeta^{(n)} \in L(\obfn^{(0)})$ for $n \geq
0$.  Since $\obfn(\cdot)$ is bounded below, $\{\obfn^{(n)}, n \geq
0\}$ evidently forms a bounded, monotone decreasing sequence and $\{
\bfbeta^{(n)}, n \geq 0\}$ forms a bounded
sequence contained wholly within the compact set $L(\obfn^{(0)})$. $\Box$\\

%% file: main_MM_thm_proof.tex
\subsection{Proof of Theorem \ref{thm:gen_penal_MM}}\label{app:genpenMM}

The assumptions stated in the theorem immediately yield that
$\obfn(\bfbeta)$ is locally Lipschitz continuous and coercive for each bounded
$\lambda > 0,$ hence (i) is satisfied.

To show (ii), we first write
\begin{eqnarray}
\nonumber
q(\bfbeta,\bfalpha;\bflam) -
p(\bfbeta;\bflam) & = &
\sum_{j=1}^p
\left[
\tilde{q}(|\beta_j|,|\alpha_j|;\bflam_j) - \tilde{p}(|\beta_j|;\bflam_j)
\right] \\
\label{q minus p}
& = & \sum_{j=1}^p
\left[
\tilde{p}(|\alpha_j|;\bflam_j) + \tilde{p}'(|\alpha_j|; \bflam_j)(|\beta_j|-|\alpha_j|)
- \tilde{p}(|\beta_j|;\bflam_j)
\right].
\end{eqnarray}
This difference is obviously equal to zero whenever $\bfbeta =
\bfalpha$. For $\bfbeta \neq \bfalpha$, we shall separately consider
the case where $\tilde{p}(r;\bflam_j)$ is linear versus nonlinear.

First, suppose that $\tilde{p}(r;\bftheta) = a_1 + a_2 r,$
where $a_1 \geq 0$ and $a_2 > 0$ and each may depend
on $\bftheta$. It then follows immediately that
\[
\tilde{p}(|\alpha_j|;\bflam_j) + \tilde{p}'(|\alpha_j|; \bflam_j)(|\beta_j|-|\alpha_j|)
- \tilde{p}(|\beta_j|;\bflam_j)
= (a_1 + a_2 |\alpha_j|) \, + \, a_2 (|\beta_j|-|\alpha_j|) \, - \,
(a_1 +  a_2 |\beta_j|) = 0.
\]
Thus, the claimed equality between
\eqref{eq:gen_pen} and \eqref{q fun} holds in this case.

Now, suppose that $\tilde{p}(r;\bftheta)$ is nonlinear in $r$.  Under
(P1), we claim that \eqref{q fun} strictly majorizes
$p(\bfbeta;\bflam)$ provided the derivative of the penalty $\tilde{p}'(\cdot,\bflam_j)$ is strictly positive. To see this, observe that concavity (e.g., see
\eqref{concavity}) implies the inequality
\[
\tilde{q}(r,s;\bftheta) - \tilde{p}(r;\bftheta)
= -1
\left[\tilde{p}(r;\bftheta) - \tilde{p}(s;\bftheta)
- \tilde{p}'(s;\bftheta)(r-s)
\right] \geq 0,
\]
with equality holding if and only if $r  = s$ and $p'(s;\bftheta)>0.$  For penalties such that their derivatives are nonnegative, i.e., $p'(s;\bftheta)\geq 0,$ we obtain the same inequality as above, with equality additionally holding for $r$ and $s$ sufficiently large. Therefore,
\[
q(\bfbeta,\bfalpha;\bflam) -
p(\bfbeta;\bflam) = \sum_{j=1}^p
\left[
\tilde{q}(|\beta_j|,|\alpha_j|;\bflam_j) - \tilde{p}(|\beta_j|;\bflam_j)
\right] \geq 0,
\]
%with equality holding if and only if $\bfbeta = \bfalpha$ and $p'(|\alpha_j|,\bflam_j)\neq 0$ for all $j$ (see also Remark \ref{deriv0}).  
and (ii) is established.

In order to establish the majorization property specified in (iii), we
begin by noting that our assumptions on $g(\bfbeta),$
$h(\bfbeta,\bfalpha)$, and $\tilde{p}(\cdot;\bftheta)$ imply that
$\sobfn(\bfbeta,\bfalpha)$ and $\psi(\bfbeta,\bfalpha) =
h(\bfbeta,\bfalpha) + q(\bfbeta,\bfalpha;\bflam) - p(\bfbeta;\bflam)$
are both continuous in $\bfbeta$ and $\bfalpha$. Our assumptions
further imply that $\psi(\bfbeta,\bfalpha) \geq 0$; if at least one of
$h(\bfbeta,\bfalpha)$ or $q(\bfbeta,\bfalpha;\bflam) -
p(\bfbeta;\bflam)$ is strictly positive for $\bfbeta \neq \bfalpha$,
then $\psi(\bfbeta,\bfalpha) > 0$ for $\bfalpha \neq \bfbeta$ and
$\psi(\bfbeta,\bfbeta) = 0$.  Therefore, the objective function
$\obfn(\bfbeta)$ is strictly majorized by $\sobfn(\bfbeta,\bfalpha)
\equiv \obfn(\bfbeta) + \psi(\bfbeta,\bfalpha).$

In order to establish the convergence of the corresponding MM
algorithm in (iii), it suffices to prove that the assumptions of the
theorem and consequent assertions established thus far are sufficient
to ensure that Conditions R1-R5 of Appendix \ref{app:MMlocal} are met,
in which case Theorem \ref{thm: Meyer} applies directly.  The result
(i), combined with the assumption that the stationary points are all
isolated, immediately establishes Condition R1; as proved above,
conditions R2 and R3 also hold. If $\psi(\bfbeta,\bfalpha) =
h(\bfbeta,\bfalpha) + q(\bfbeta,\bfalpha;\bflam) - p(\bfbeta;\bflam)$
is continuous in $\bfalpha$ and $\bfbeta$ and locally Lipschitz
continuous in $\bfbeta$ near $\bfalpha$, then (i) implies that R4 also
holds. By assumption, $h(\bfbeta,\bfalpha)$ is continuous in
$\bfalpha$ and continuously differentiable in $\bfbeta$, hence locally
Lipschitz in $\bfbeta$. Continuity of $q(\bfbeta,\bfalpha;\bflam) -
p(\bfbeta;\bflam)$ in both $\bfalpha$ and $\bfbeta$ is also immediate.
Hence, R4 holds provided that $q(\bfbeta,\bfalpha;\bflam) -
p(\bfbeta;\bflam)$ is locally Lipschitz continuous in $\bfbeta$ near
$\bfalpha$.  To see that this is the case, we note that \eqref{q minus
  p} is a linear combination of functions in $\beta_j$ of the form
$\tilde{p}'(|\alpha_j|; \bflam_j)|\beta_j|-
\tilde{p}(|\beta_j|;\bflam_j),$ where $| \cdot |$ and
$-\tilde{p}(\cdot;\bflam)$ are both convex, hence locally
Lipschitz. Since both the sum and composition of two locally Lipschitz
functions are locally Lipschitz, the result now follows. Finally, R5
is ensured by R1-R4 and the condition in (iii) that
$\sobfn(\bfbeta,\bfalpha)$ is uniquely minimized in $\bfbeta$ for each
$\bfalpha$.

%% file: thm_gen_MIST_proof.tex
\subsection{Proof of Theorem \ref{thm: gen_MIST}}
\label{app:genMIST}

Under the stated conditions and for any bounded $\bfalpha,$
$m(\bfbeta) = g(\bfbeta) + h(\bfbeta,\bfalpha) + \lambda \epsilon \|
\bfbeta \|^2$ is strictly convex with a Lipschitz continuous
derivative of order $L^{-1} > 0$; in addition, $\sum_{j=1}^p
\tilde{p}'(|\alpha_j|; \bflam_j) |\beta_j|$ is also convex in
$\bfbeta$. Hence, for each bounded $\bfalpha$ there exists a unique
solution $\bfbeta^* = \bfbeta^*(\bfalpha)$ when minimizing \eqref{eq:
  MIST object 1}.

In the notation of \cite{cw05}, we may identify the Hilbert
space ${\cal H}$ with $\Real^p$, $f_2(\bfbeta)$ with $m(\bfbeta)$ and
$f_1(\bfbeta)$ with $\sum_{j=1}^p \tilde{p}'(|\alpha_j|; \bflam_j)
|\beta_j|$. The assumptions of the theorem ensure that
the regularity conditions of Proposition 3.1 and Theorem 3.4
of \cite{cw05} are met. In particular,
because $m(\bfbeta)$ is coercive and strictly convex,
Proposition 3.1 guarantees the existence of a unique solution
to
\[
\min_{\bfbeta \in \Real^p}
f_1(\bfbeta) + f_2(\bfbeta)
\]
as well as provides the relevant fixed point mapping; Theorem 3.4
establishes the weak convergence of the corresponding iterative scheme
to this unique solution. Since weak convergence is equivalent to strong
convergence in a finite dimensional Hilbert space, such results imply
componentwise convergence of the resulting iteration
sequence to $\bfbeta^*$.

Both Proposition 3.1 and Theorem 3.4 of \cite{cw05} rely
on the gradient of $f_2(\bfbeta)$ and the so-called ``proximity operator''
of $f_1(\bfbeta)$.  Example 2.20 in \cite{cw05} shows
that the proximity operator for $f_1(\bfbeta) = \sum_{j=1}^p
\tilde{p}'(|\alpha_j|; \bflam_j) |\beta_j|$ is exactly $S(\cdot
;\bftau).$ The algorithm summarized in the statement of the theorem is
therefore observed to be a specific instance of that described in the
Theorem 3.4 with (in their notation) $a_n = b_n = 0$ and $\lambda_n =
1$ for every $n$.

\citet[][Theorem 4.5]{hyz08} undertake a detailed study of the
proposed algorithm for the special case of a convex, differentiable
$f_2(\bfbeta)$ and $f_1(\bfbeta) = \sum_{j=1}^p |\beta_j|$.  In this
case, they prove that the algorithm converges in a finite number of
iterations.

A minor extension of their arguments may be used to
establish the same result for $f_1(\bfbeta) = \sum_{j=1}^p
\tilde{p}'(|\alpha_j|; \bflam_j) |\beta_j|,$ provided that
$\tilde{p}'(|\alpha_j|; \bflam_j) \in [0,\infty)$ for each $j$.

%% file: glm_thm_proof.tex
%==================================================================
%==================================================================
\subsection{Proof of Theorem \ref{prop:sur:glm}}
\label{app:surglm}

%%%%%%%%%%%%%%%%%%%%%%%%%%%%%%%%%%%%%%%%%%%%%%%%%%%%%%%%%%%%%%%%%%%%%%%%%%%%
%\hspace*{0.0in}

% \textbf{EDS: Is it ok to leave everything as is for (3.) below, with
%   "differentiably in $\beta_j$ only if $\beta_j\neq 0$" etc, or do we
%   need to make this more rigorous with the subdifferentials?
%   Actually, is this part even needed at all?}

%\textbf{EDS: do we really need to include proofs for 1 and 2}\\
%
%\textbf{RLS: there should be a brief discussion of these conditions
%and why these hold; full,
%detailed proof need not be given.}\\
%%%%%%%%%%%%%%%%%%%%%%%%%%%%%%%%%%%%%%%%%%%%%%%%%%%%%%%%%%%%%%%%%%%%%%%%%%%%

To establish (1.), note that the choice of $h(\tbfbeta,\tbfalpha)$
in{~\eqref{h glm}} with appropriate $\varpi$ guarantees majorization
of $-\ell(\tbfbeta)$ provided $\nabla^2(-\ell(\tbfbeta))$ can be
bounded \citep[e.g., ][Ch 6]{optim}.  Penalties of form
\eqref{eq:gen_pen} satisfying assumption (P1) can be linearly
majorized so that \eqref{eq: MIST_glm} majorizes
$\xi_{glm}(\tilde{\bfbeta}).$ For (2.), $-\ell(\tbfbeta)$ is indeed
strictly convex and coercive, with $h(\tbfbeta,\tbfalpha)\geq 0$
continuous in both $\tbfbeta$ and $\tbfalpha$ and continuously
differentiable in $\tbfbeta$ for each $\tbfalpha,$ with
$h(\tbfbeta,\tbfalpha) = 0$ when $\tbfbeta = \tbfalpha$.  We provide a
more detailed proof of (3.) below.  Let $\zeta = 2\varpi^{-1}.$ Note
that the surrogate $\sobfn(\tbfbeta,\tbfalpha)$ is differentiable in
$\beta_j$ only if $\beta_j \neq 0$.  Assuming $\beta_j \neq 0$ for
$j\neq 0$ and excluding irrelevant constants,

\begin{equation}\label{genderiv}
	\frac{\partial \sobfn_{glm}(\tilde{\bfbeta};\tilde{\bfalpha}) }{\partial \beta_j} =
		- \left[ \nabla \ell(\tilde{\bfalpha}) \right]_j + 
\zeta \beta_j -  \zeta \alpha_j
+ \tau_j\textrm{sign}(\beta_j) + 2\lambda\epsilon\beta_j.
\end{equation}

Setting{~\eqref{genderiv}} equal to zero implies 
$$
\beta_j = \left\{ \begin{array}{ll}
 \frac{1}{\zeta + 2\lambda\epsilon}
\left( \left[ \nabla \ell(\tilde{\bfalpha}) \right]_j + 
\zeta \alpha_j - \tau_j \right) & \beta_j>0 \\
 \frac{1}{\zeta + 2\lambda\epsilon}
\left( \left[ \nabla \ell(\tilde{\bfalpha}) \right]_j + 
\zeta \alpha_j + \tau_j \right) & \beta_j<0 \\
\end{array} \right.
.
$$

For sign consistency, we must impose that $
\frac{1}{\zeta + 2\lambda\epsilon}\left( \left[ \nabla
    \ell(\tilde{\bfalpha}) \right]_j + \zeta \alpha_j \right) > \tau_j$
when $\beta_j>0$ and $ \frac{1}{\zeta + 2\lambda\epsilon}\left( \left[
    \nabla \ell(\tilde{\bfalpha}) \right]_j + \zeta \alpha_j \right) < -
\tau_j $ when $\beta_j<0.$

When $ \left| \frac{1}{\zeta+2\lambda\epsilon}\left( \left[ \nabla
      \ell(\tilde{\bfalpha}) \right]_j + \zeta \alpha_j \right) \right|
\leq \tau_j,$ we set $\beta_j = 0$.  In summary,
%\begin{equation}
$$	\beta_j^{\ast} = \frac{1}{\zeta+2\lambda\epsilon}s
\left( \left[ \nabla \ell(\tilde{\bfalpha}) \right]_j + 
\zeta \alpha_j, \tau_j \right),$$
%\end{equation}
from which the first part of \eqref{beta.star} directly follows for
$j\in\{1,\ldots,p\}$.  We do not penalize the intercept, thus
 %\begin{equation}%\label{genderiv0}
	$$\frac{\partial 
\sobfn_{glm}(\tilde{\bfbeta};\tilde{\bfalpha}) }{\partial \beta_0} =
        - \left[ \nabla \ell(\tilde{\bfalpha}) \right]_0 + \zeta \beta_0 -
        \zeta \alpha_0$$
	%\end{equation}
so that $\beta_0^{\ast} = (\left[ \nabla \ell(\tilde{\bfalpha}) \right]_0 
+  \zeta \alpha_0)/\zeta.$

Furthermore, take $\tilde{\bfbeta} + \tilde{\bfh}$ for any $\tilde{\bfbeta} \in \mathcal{R}^{p+1}$ and 
$\tilde{\bfh} = (\kappa_0,\bfh^T)^T \in \mathcal{R}^{p+1}$ is arbitrary.  
Then, following arguments similar to those in \citet[Prop. 2.1]{daub},
\begin{eqnarray}
\nonumber
\xi^{SUR}_{glm}( \tilde{\bfbeta} + \tilde{\bfh}, \tilde{\bfalpha}   ) &=&  -\ell(\tilde{\bfalpha}) - \nabla \ell(\tilde{\bfalpha})'(\tilde{\bfbeta}+\tilde{\bfh}-\tilde{\bfalpha})  + \frac{\zeta}{2}(\tilde{\bfbeta}+\tilde{\bfh} - \tilde{\bfalpha})'(\tilde{\bfbeta}+\tilde{\bfh} - \tilde{\bfalpha)}\\
\nonumber
&& + \sum_{j=1}^p (\tau_j|\beta_j+\kappa_j| + \gamma_j + 
\lambda\epsilon(\beta_j+\kappa_j)^2) \\
%\label{min.sur:glm}
\nonumber
&=& \xi^{SUR}_{glm}( \tilde{\bfbeta}, \tilde{\bfalpha}  )+ (\frac{\zeta}{2}+\lambda\epsilon)\bfh'\bfh + \frac{\zeta}{2}\kappa_0^2 + \kappa_0(\zeta \beta_0 - \zeta \alpha_0 - [\nabla\ell(\tilde{\bfalpha})]_0)\\
%\bfh'(\bfbeta-\bfalpha - \bfX'(\bfy-\bfX\bfalpha))\\
\nonumber
&&+ \sum_{j=1}^p \left[ \tau_j(|\beta_j+\kappa_j| - |\beta_j|) + \kappa_j((d+2\lambda\epsilon)\beta_j-\zeta \alpha_j - [\nabla\ell(\tbfalpha)]_j) \right].
\end{eqnarray}
Consider
$\tilde{\bfbeta}=\tilde{\bfbeta}^{\ast}\equiv[\beta_0^{\ast},\bfbeta^{\ast
  T}]^T$ where $\tilde{\bfbeta}^\ast$ defined in{~\eqref{beta.star}},
and define sets $\mathcal{J}=\{1,2,\ldots,p\},$
$\mathcal{J}_0=\{j\in\mathcal{J}: \beta_j^\ast=0\}$ and
$\mathcal{J}_1=\mathcal{J}\backslash\mathcal{J}_0$.  Noting that
$\beta_j^\ast$ satisfies $(\zeta+2\lambda\epsilon)\beta_j^\ast - \zeta
\alpha_j - [\nabla\ell(\bfalpha)]_j =
-\tau_j\textrm{sign}(\beta_j^\ast)$ for $j\in\mathcal{J}_1$, and
noting that $\zeta \beta_0^{\ast} -\zeta \alpha_0 - [\nabla
\ell(\tilde{\bfalpha})]_0 = 0,$ we have
\begin{eqnarray}
\nonumber
\xi^{SUR}_{glm}( \tilde{\bfbeta}^\ast + \tilde{\bfh}, \tilde{\bfalpha}  ) - \xi^{SUR}_{glm}( \tilde{\bfbeta}^\ast, \tilde{\bfalpha}  ) &=& (\frac{\zeta}{2}+\lambda\epsilon)\bfh'\bfh + \frac{\zeta}{2}\kappa_0^2 + \kappa_0(\zeta \beta_0^{\ast} -\zeta \alpha_0 - [\nabla \ell(\tilde{\tbfalpha})]_0)  \\
\nonumber
&&  + \sum_{j=1}^p \tau_j(|\beta_j^\ast+\kappa_j| - |\beta_j^\ast|) + \kappa_j((\zeta+2\lambda\epsilon)\beta_j^\ast-\zeta \alpha_j - [\nabla\ell(\tbfalpha)]_j)\\
\nonumber
&=& (\frac{\zeta}{2}+\lambda\epsilon)\bfh'\bfh + \frac{\zeta}{2}\kappa_0^2 + \sum_{j\in\mathcal{J}_0}\left[\tau_j|\kappa_j| - \kappa_j(\zeta \alpha_j + [\nabla\ell(\bfalpha)]_j) \right] \\
\nonumber
&& + \sum_{j\in\mathcal{J}_1}\left[\tau_j(|\beta_j^\ast+\kappa_j| - |\beta_j^\ast|)- \kappa_j\tau_j\textrm{sign}(\beta_j^\ast)\right].
\end{eqnarray}
For $j \in \mathcal{J}_0,$ $|\zeta \alpha_j + [\nabla\ell(\tbfalpha)]_j| \leq \tau_j,$ so that $\tau_j|\kappa_j| - \kappa_j(\zeta \alpha_j + [\nabla\ell(\tbfalpha)]_j) \geq 0.$  For $j \in \mathcal{J}_1,$ there are two cases, corresponding to the sign of $\beta_j^\ast.$ First consider $\beta_j^\ast>0$, then
\begin{eqnarray}
\nonumber
\tau_j(|\beta^\ast_j+\kappa_j| - |\beta^\ast_j|) -\kappa_j\tau_j\textrm{sign}(\beta_j^\ast) %&=& \tau_j(|\beta_j+\kappa_j| - \beta_j -\kappa_j\tau_j\\
&=& \tau_j(|\beta^\ast_j+\kappa_j| - (\beta^\ast_j+\kappa_j)) \geq 0.
\end{eqnarray}
If  $\beta_j^\ast<0$, then
\begin{eqnarray}
\nonumber
\tau_j(|\beta_j^\ast+\kappa_j| - |\beta_j^\ast|) -\kappa_j\tau_j\textrm{sign}(\beta_j^\ast) %&=& \tau_j(|\beta_j+\kappa_j| + \beta_j +\kappa_j\tau_j\\
&=& \tau_j(|\beta_j^\ast+\kappa_j| + (\beta_j^\ast+\kappa_j)) \geq 0.
\end{eqnarray}
Thus, $\xi^{SUR}_{glm}( \tilde{\bfbeta}^\ast + \tilde{\bfh}, \tilde{\bfalpha}  ) - \xi^{SUR}_{glm}( \tilde{\bfbeta}^\ast, \tilde{\bfalpha}  ) \geq (\frac{\zeta}{2}+\lambda\epsilon)\bfh'\bfh + \frac{\zeta}{2}\kappa_0^2 \geq \frac{\zeta}{2}\tilde{\bfh}'\tilde{\bfh},$ since $\lambda\epsilon \geq 0,$ hence guaranteeing a unique minimum, and proving the proposition.
$\Box$\\